\documentclass[aps,prx,amsmath,amssymb,reprint,longbibliography]{revtex4-2}
\usepackage[english]{babel}
\usepackage{mathtools}
\usepackage{graphicx}
\usepackage{mathrsfs}
\usepackage{amssymb}
\usepackage{enumitem}

\usepackage{xcolor}
\usepackage[colorlinks=true,citecolor=blue,urlcolor=blue]{hyperref}
\usepackage{amsmath}
\usepackage{bm}
\usepackage{physics}

\renewcommand{\selectlanguage}[1]{}
\usepackage[normalem]{ulem}
\begin{document}
\newcommand{\NL}[1]{\textcolor{blue}{#1}}
\newcommand{\commentNikita}[1]{\textcolor{blue}{\textbf{Nikita:} \textit{#1}}}
\title{Quantum correlated steady states under competing collective and individual decay}

\author{Nikita Leppenen}
\author{Ephraim Shahmoon}
\affiliation{Department of Chemical \& Biological Physics, Weizmann Institute of Science, Rehovot 7610001, Israel}

\begin{abstract}
%
%
Collective dissipation can generate useful quantum correlations, while ubiquitous individual decay destroys them. We study the interplay between these two competing processes considering a driven system of many spins (``atoms") undergoing both collective and individual dissipation (``radiation").
In steady state and depending on drive, we find that the system exhibits a first-order phase transition and quantum bistability: its quantum state is a mixture of two many-body states associated with the two competing decay processes. Accordingly, one of these states closely resembles a correlated ``coherently radiating spin state" (CRSS) --- the solution of purely collective dissipation --- exhibiting spin-squeezing entanglement. We predict dynamical switching between the two stable states, manifest as many-body quantum jumps in the various observables of spin and radiation. Macroscopically, the switching rate tends to vanish and the system can reside in a correlated CRSS for long times.
This reveals how correlated dissipative physics emerges at the presence of decorrelating individual decay, opening a path for unlocking collective dissipation phenomena in realistic quantum platforms and applications. We discuss consequences for experiments in collective radiation.
\end{abstract}

\maketitle

\section{Introduction}

Collective dissipation is formed when two or more systems are damped to a common reservoir. In many-body quantum systems, this reservoir-mediated interaction can lead to robust generation of entanglement, with applications in quantum information, simulation and metrology~\cite{reiter_cooperative_2020,kraus_preparation_2008, ma_quantum_2011}.
In practice, however, any realistic system exhibits dissipation also at the individual-constituent level. Such individual decay destroys the entanglement, hence posing a main limitation in current quantum platforms and processors~\cite{reiter_dissipative_2017,suzuki_quantum_2022,hoke_measurement-induced_2023}. The question is then whether and how entangled states generated by collective dissipation survive at the presence of individual decay. More generally, this amounts to the interplay between collective and individual dissipation: in particular, what are the quantum many-body states and phases that are stabilized by these two competing decay processes in steady state? Here, we study this problem considering systems comprised of many effective spins relevant to platforms of quantum science.

\begin{figure}[t!]
    \centering
    \includegraphics[width=\columnwidth]{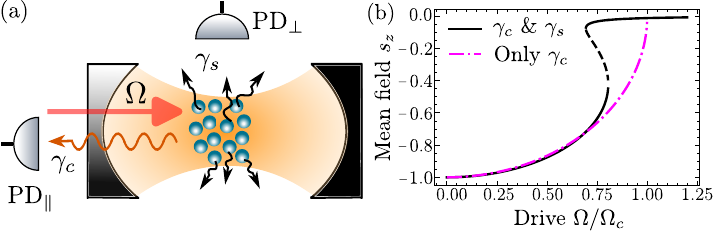}
    \caption{(a) Quantum optical realization of the model, Eq.~(\ref{Eq:ME}). Atoms trapped inside an optical cavity and coherently driven by a laser amplitude $\Omega$. The $N$ atoms decay collectively through the cavity mirror with rate $\gamma_c$ and individually to off-axis modes at rate $\gamma_s$. The photon detectors PD$_{\parallel}$ and PD$_{\perp}$ give access to the collective atomic observables $\hat{S}_{\pm}$ (and hence $\hat{S}_{x,y}$, $\hat{S}_+\hat{S}_-$) and the ``magnetization"  (population inversion) $\hat{S}_z$, respectively.
    (b) Mean-field solution of the population inversion per atom $s_z$ from  Eq.~\eqref{eq:gammaneq0} (solid and dashed curve for stable and unstable solutions), with $\Omega_c=\Gamma/4$ and taking $\Gamma/\gamma = \gamma_c(N-1)/(\gamma_c+\gamma_s)=15$. Dicke case ($\gamma_s=0$) is plotted for reference (dash-dotted).
    }
    \label{fig:sketch}
\end{figure}

A paradigmatic case of collective dissipation arises naturally in the radiation from many atom-like emitters (``atoms"). The atoms, considered as two-level pseudo-spins, interact via the multiple scattering of reservoir photons, as occurs in various systems and applications~\cite{gross_superradiance_1982,meiser_prospects_2009,henriet_critical_2019,chang_controlling_2004,ostermann_breakdown_2023,norcia_superradiance_2016,kaluzny_observation_1983,angerer_superradiant_2018,angerer_superradiant_2018,kersten_self-induced_2024,bohnet_steady-state_2012,lin_chapter_2012,pellegrino_observation_2014,kersten_triggered_2023,grimes_direct_2017,ferioli_non-equilibrium_2023,masson_universality_2022,ishizaki_quantum_2012,gonzalez-tudela_deterministic_2015,wang_supercorrelated_2020,rivero_quantum_2023,hotter_cavity_2023,song_dissipation_2024}. In its purest form, the essence of such collective radiation is captured by assuming that all atoms are uniformly coupled to the relevant photon modes, as envisioned for a cavity (Fig.~\ref{fig:sketch}). The many atoms then appear indistinguishable to the photon field and radiate as a single collective dipole, forming the Dicke model of purely collective dissipation~\cite{dicke_coherence_1954,gross_superradiance_1982}. Steady state is reached when dissipation is balanced by a coherent drive (e.g. a laser) exhibiting a continuous dissipative phase transition at a critical drive~\cite{drummond_volterra_1978,carmichael_analytical_1980,kessler_dissipative_2012,sanchez_munoz_symmetries_2019,barberena_driven-dissipative_2019}. Below the transition, the system was recently found to be described by the ``coherently radiating spin state" (CRSS)~\cite{somech_quantum_2024}: a pure many-body state, which exhibits metrologically useful quantum entanglement of the atoms (spins)~\cite{gonzalez-tudela_mesoscopic_2013,lee_dissipative_2014,qu_spin_2019} while radiating classical coherent-state light~\cite{somech_heisenberg-langevin_2023}.
A full analytical characterization of purely collective dissipation is therefore available in terms of CRSS.

Nevertheless, the Dicke indistinguishability symmetry between the atoms, which underlies purely collective dissipation, never really occurs in typical quantum platforms such as atomic arrays~\cite{bloch_ultracold_2005,barredo_atom-by-atom_2016,endres_atom-by-atom_2016,rui_subradiant_2020,bettles_enhanced_2016,shahmoon_cooperative_2017,asenjo-garcia_exponential_2017,grankin_free-space_2018,rui_subradiant_2020,cidrim_photon_2020,parmee_bistable_2021,robicheaux_beyond_2021,fernandez-fernandez_tunable_2022,pedersen_quantum_2023,solomons_universal_2023,yan_superradiant_2023} and ensembles ~\cite{hammerer_quantum_2010,bromley_collective_2016,guerin_subradiance_2016,ferioli_non-equilibrium_2023} due to the non-uniform  coupling of atoms to multiple photon modes \cite{gross_superradiance_1982,chang_controlling_2004,meiser_prospects_2009,olmos_steady-state_2014,henriet_critical_2019,qu_spin_2019,ostermann_breakdown_2023,goncalves_driven-dissipative_2024, agarwal_directional_2024}. Therefore, even when it is possible to identify a common photon mode that mediates collective dissipation, it will be always accompanied by effective individual dissipation to multiple other modes. This gives a concrete perspective to the need for studying the interplay between collective and individual dissipation.

To this end, we consider a simple model that exhibits this interplay: coherently driven atoms (spins) undergoing purely collective dissipation in parallel to individual single-atom decay. This has important advantages: (i) the two competing dissipation processes are well-defined, allowing to reveal the essence of their interplay in a clear and systematic fashion; (ii) while simple, this model corresponds to a realistic description of the most common setup of collective radiation --- that of atoms in a cavity [Fig.~\ref{fig:sketch}(a)]~\cite{kaluzny_observation_1983,norcia_superradiance_2016,angerer_superradiant_2018,rivero_quantum_2023,kersten_self-induced_2024}. 
This allows to relate our results to current platforms, providing new insights and predictions.


We perform a full quantum analysis of the problem in steady-state using a combination of numerical and analytical approaches, going well beyond mean-field \cite{drummond_volterra_1978,carmichael_analytical_1980} and transient-times~\cite{tucker_facilitating_2020,masson_many-body_2020,masson_universality_2022} treatments. We find the following:

\begin{itemize}[leftmargin=*]
\item The interplay between collective and individual dissipation leads to the emergence of two phases, one dominated by correlated physics and the other of largely independent spins. Each phase is fully characterized by a distinct quantum many-body state which we find: The first is close to an entangled CRSS --- the solution of purely collective dissipation, while the second approaches a separable mixed state --- the solution of individual dissipation.
Accordingly, the two corresponding states occupy distinct regions in the Hilbert space, spanned mostly by large or small total angular momenta states, respectively. Several features are revealed as follows.
  \item \emph{First-order dissipative phase transition (Sec.~\ref{sec:1stPT}):} As a function of the drive $\Omega$, a first-order transition between the correlated and individual phases is exhibited in the variety of physical observables and correlations of both spins and radiation. 
  \item \emph{Quantum bistability (Sec.~\ref{sec:bistability}) and switching dynamics (Sec.~\ref{sec:SS_dyn}):} There exists a region of drive values in which the system is bistable, i.e. the steady-state density matrix is given by a mixture, $\rho_s=a_+\rho_+ + a_-\rho_-$, of the two many-body states corresponding to the correlated ($\rho_+$) and individual ($\rho_-$) phases. While the mixture $\rho_s$ describes the average state, the actual dynamics is that of switching between the states $\rho_{\pm}$ via many-body quantum jumps. Accordingly, dynamical switching between the values corresponding to the correlated and individual phases is exhibited in all observables. The switching rate vanishes with system size, hence the system can reside in the correlated state $\rho_+$ for long times.
  \item \emph{Emergence of correlations (Sec.~\ref{sec:CRSS}):} The correlated state $\rho_+$ is shown to resemble the CRSS associated with pure collective dissipation. The Optimum of spin-squeezing entanglement is derived analytically and coherent-state radiation is found.
  \item \emph{Consequences on observations (Sec.~\ref{sec:exp}):} Correlated Dicke-like physics is observable either in transient or in long times via $\rho_+$. At long times, the preparation of either of the quasi-stationary states $\rho_+$ or $\rho_-$ is controlled by the initial total angular momentum.
\end{itemize}

Ultimately, our results reveal how correlated physics associated with collective dissipation emerges at the presence of decorrelating individual decay. Essentially, this occurs via the formation of a quantum bistability containing a correlated CRSS-like state as one of its stable phases. This provides a generic mechanism for the generation of steady-state entanglement in collective dissipation problems beyond the Dicke indistinguishablity symmetry, in relation to relevant quantum platforms.



\section{Model}
\label{sec:model}
We consider $N$ identical two-level atoms (spins) that are coherently driven and undergoing both collective and individual decay. These three processes are described by the master equation for the many-atom density matrix $\rho$ which includes the three respective terms,
\begin{multline}\label{Eq:ME}
	\dot{\rho} = -i [2\Omega\hat{S}_x,\rho]+\frac{\gamma_{\mathrm{c}}}{2}\qty(2\hat{S}_-\rho \hat{S}_+-\rho \hat{S}_+ \hat{S}_- - \hat{S}_+ \hat{S}_- \rho)\\+\frac{\gamma_s}{2}\sum_{n = 1}^N\qty(2\hat{\sigma}_n \rho \hat{\sigma}_n^\dagger -\rho \hat{\sigma}_n^\dagger \hat{\sigma}_n -\hat{\sigma}_n^\dagger \hat{\sigma}_n\rho) \equiv -{\cal L}\rho.
\end{multline}
Here $\Omega$ is the amplitude of the resonant coherent drive (Rabi frequency), while $\gamma_c$ and $\gamma_s$ denote the decay rates to the collective and individual dissipation channels, respectively, with $\hat{S}_- = \hat{S}_+^\dagger = \sum_{n = 1}^N \hat{\sigma}_n$  being the atomic collective-spin lowering operator, $\hat{\sigma}_n$  the pseudo-spin lowering operator of a single atom $n \in \{1,..., N\}$, and similarly $\hat{S}_{x,y,z} = 1/2\sum_{n = n}^N \hat{\sigma}_n^{x,y,z}$.

A natural physical realization of Eq. (\ref{Eq:ME}) is the radiation of laser-driven atoms in a cavity. In this case, the second term captures collective radiation of the atoms via the cavity mirrors at a Purcell-enhanced rate $\gamma_c$: it is obtained by assuming all atoms are identically coupled to the cavity mode (e.g. laser-trapped at its antinodes~\cite{norcia_superradiance_2016, song_dissipation_2024}), and adiabatically eliminating the fast-decaying cavity mode (Appendix~\ref{app:HL_eqs}). The third term accounts for additional decay of the atoms to off-axis modes outside the cavity: assuming inter-atomic separations exceeding the optical wavelength, the atoms appear distinguishable to these modes. This effectively leads to individual decay at the free-space spontaneous emission rate $\gamma_s$, described by individual-atom operators $\hat{\sigma}_n$.

Within this setup the atomic observables can be related to detection of scattered light~\cite{ferioli_non-equilibrium_2023}. First, the amplitude of the field $\hat{E}_c$ radiated via the cavity mirrors,  as can be measured by photodetector PD${}_{\parallel}$ in Fig.~\ref{fig:sketch}, is proportional to the collective dipole amplitude $\hat{S}_-$~\cite{somech_heisenberg-langevin_2023, goncalves_driven-dissipative_2024} (see also Appendix~\ref{app:HL_eqs}). Then, homodyne detection of the field's quadratures, $\hat{E}_c \pm \hat{E}^{\dag}_c$, gives access to collective spin observables $\hat{S}_x$ and $\hat{S}_y$, while the direct photocurrent yields the collective intensity $\hat{S}_+\hat{S}_-$.
Conversely, the intensity of emission to off-cavity-axis, measured by photodetector PD${}_\perp$ (Figure~\ref{fig:sketch}), is proportional to the individual-atom population $(1+\hat{\sigma}_n^z)/2$ summed over the atoms, yielding a measurement of the ``magentization" $\hat{S}_z$. This allows access to all of the collective atomic observables $\hat{S}_{x,y,z}$ in addition to the radiation field $\hat{E}_c$.


\section{First-order phase transition}
\label{sec:1stPT}
\subsection{Mean-field analysis}
\label{sec:1spPT_mf}
Before we turn to a full quantum treatment, it is instructive to revisit the mean-field analysis of Eq.~\eqref{Eq:ME}. At long times, the individual decay decorrelates different atoms, which are nonetheless statistically equivalent. This motivates to use an individual factorization approximation with identical mean-field values for different atoms~\cite{drummond_volterra_1978}: $\expval{\hat{\sigma}_n}_{\rm MF} = s$, $\expval{\hat{\sigma}_n^z}_{\rm MF} = s_z$ and {$\expval{\hat{\sigma}_n^\dagger \hat{\sigma}_m}_{\rm MF}  = \abs{s}^2$} for $n \neq m$.  We analyze the resulting mean-field dynamical equations in Appendix~\ref{app:MF}, finding their steady state as the solution of the cubic equation
\begin{eqnarray}\label{eq:gammaneq0}
&&(1+s_z)(\gamma-s_z \Gamma)^2+8 s_z \Omega^2 = 0,
\nonumber \\
&& \gamma = \gamma_s+\gamma_c,
\quad
\Gamma = (N-1)\gamma_c,
\end{eqnarray}
with $\gamma$ and $\Gamma$ forming the relevant mean-field parameters.

For $\Gamma>8\gamma$ there always exists a region of $\Omega$ values wherein Eq.~\eqref{eq:gammaneq0} has two stable (and one unstable) solutions. This bistable region is seen in Fig.~\ref{fig:sketch}(b) by the numerical solution of Eq. (\ref{eq:gammaneq0}). Importantly, the condition $\Gamma>8\gamma$ for bistability is easily met in current cavity QED experiments employing either macroscopic atomic ensembles with $N \sim 10^5$~\cite{norcia_superradiance_2016,rivero_quantum_2023,song_dissipation_2024} or mesoscopic tweezer arrays of $N\sim 18$ atoms trapped inside high-cooperativity cavities $\gamma_{\rm c}/\gamma_{s} \geq 10$~\cite{yan_superradiant_2023}. In the regime $\Gamma\gg \gamma$ where collective effects are dominant, we also solved Eq.~\eqref{eq:gammaneq0} analytically up to third order in $\gamma/\Gamma\ll 1$ (Appendix~\ref{app:MF}). The lowest order solution recovers the result~\cite{carmichael_analytical_1980}
\begin{equation}\label{eq:sz_an}
	s_z^{(^a_b)} = -\frac{1}{2}\mp \frac{1}{2}\sqrt{1-\frac{2\Omega^2}{\Omega_c^2}}, \quad s_z^{(c)} = 0,
\quad
\Omega_c = \frac{\Gamma}{4},
\end{equation}
exhibiting the bistability region $0<\Omega <\Omega_c/\sqrt{2}$.

Notably, this bistable behavior, typical of first-order transitions, is in contrast to the Dicke case $\gamma_s=0$: there, mean-field is performed in collective variables such that correlations are retained, obtaining a second-order transition, $s_z = -\sqrt{1-(\Omega/\Omega_c)^2}$ \cite{drummond_volterra_1978,lee_dissipative_2014,somech_heisenberg-langevin_2023}, as seen in Fig.~\ref{fig:sketch}(b).
However, for $\gamma_s\neq 0$ at long times, correlations may break and Eq. \eqref{eq:sz_an} [or~\eqref{eq:gammaneq0}] forms the solution.

\subsection{Full quantum solution}

\begin{figure}[t!]
    \includegraphics[width=1\columnwidth]{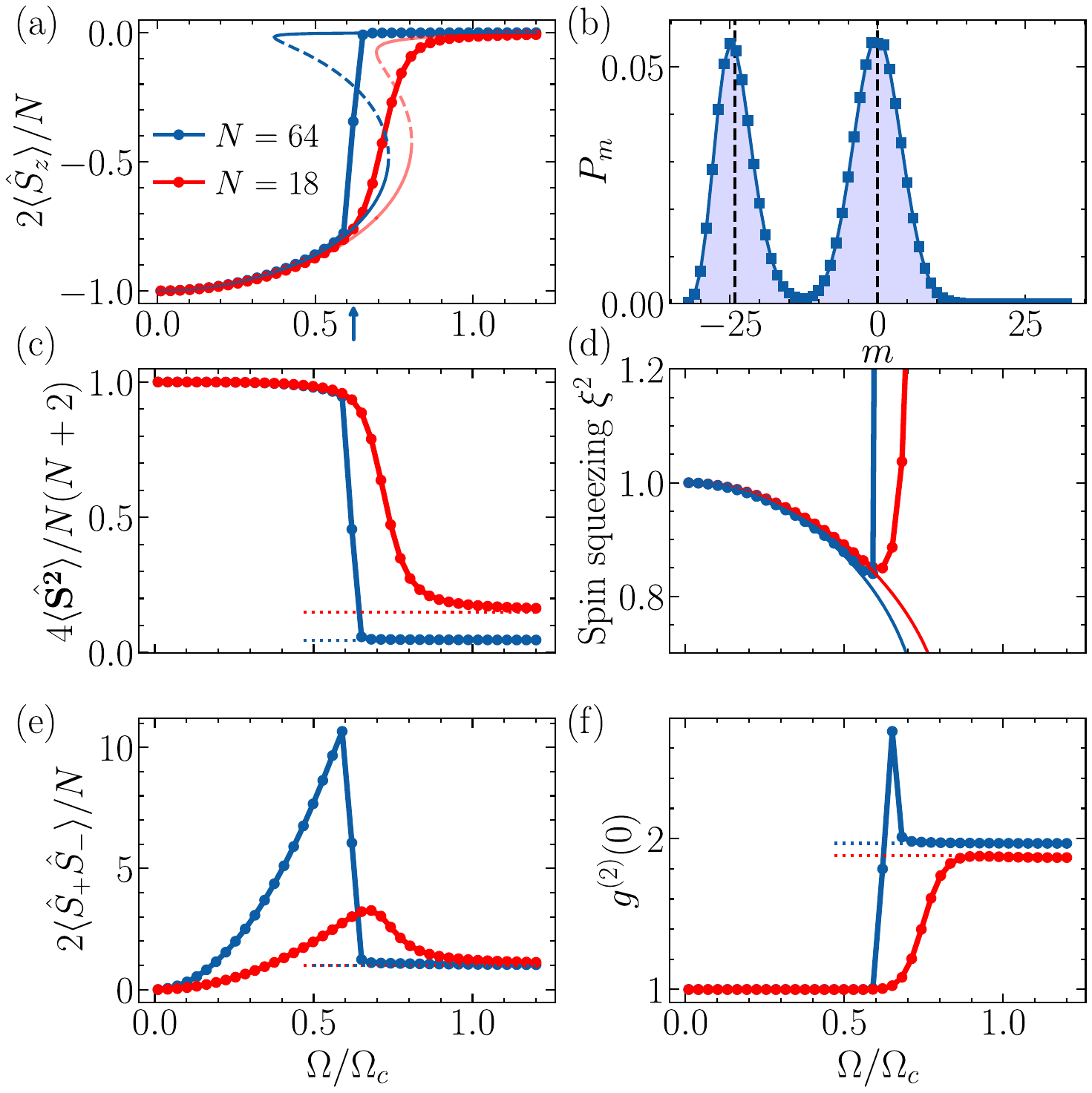}
    \caption{First-order dissipative phase transition revealed in the numerical solution of the quantum master equation~\eqref{Eq:ME} in steady state, for $N = 18$ (red curves) and $N=64$ (blue curves), and for fixed decay rates $\gamma_{c} = 10\gamma_{s}$. The critical point is estimated at $\Omega_{\mathrm{PT}}\approx 0.61\Omega_c$ for $N\gg 1$ (text and Appendix~\ref{app:PTpoint}). (a) Average population inversion (``magnetization") $\langle \hat{S}_z\rangle$ as a function of the drive $\Omega$ (solid curves with dots) compared with the mean-field solution of Eq.~\eqref{eq:gammaneq0} (thin-line curves). (b) Probability distribution of the magnetization eigenvalues $m$ at $ \Omega_{\rm PT}$ for $N = 64$ exhibits a bimodal form centered at the two stable mean-field solutions. (c) Average total angular momentum $\hat{\bm S}^2$ reveals a transition between collective- and individual-physics phases (large and small total angular momenta, respectively; dotted lines mark results of a totally mixed product state of purely individual physics, Appendix~\ref{app:mixed}).  (d) Spin squeezing parameter shows existence of quantum correlations, $\xi^2<1$, at the collective phase. Thin solid curves display analytical results from Eq. (\ref{eq:xi}). (e,f) Average intensity $\langle\hat{S}_+\hat{S}_-\rangle$ and second-order coherence $g^{(2)}(0)$ of collectively radiated light.}
    \label{fig:res1}
\end{figure}

The full solution of Eq.~\eqref{Eq:ME} can be evaluated efficiently at complexity ${\cal O}(N^3)$ by exploiting the statistical equivalence between atoms whose individual decay coefficients $\gamma_s$ are identical~\cite{xu_simulating_2013,damanet_cooperative_2016,shammah_superradiance_2017,zhang_monte-carlo_2018} (see~\cite{roberts_exact_2023} for an alternative approach). We used PIQS~\cite{shammah_open_2018} implemented in QuTip~\cite{johansson_qutip_2013} where such an algorithm is realized to calculate numerically the density matrix and atomic observables.

In Fig.~\ref{fig:res1}(a), we compare the steady-state population inversion (``magnetization") $\langle\hat{S}_z\rangle$ from the exact solution of (\ref{Eq:ME}) for different atom numbers $N$, to the mean-field result $N s_z/2$ with $s_z$ from Eq.~\eqref{eq:gammaneq0}. We observe that while the two solutions agree outside the bistability region predicted by mean-field, the comparison within this region is more subtle since the steady-state $\rho_s$ of Eq.~\eqref{Eq:ME} is unique~\cite{nigro_uniqueness_2019}. To capture the bistability effect, we plot in Fig.~\ref{fig:res1}(b) the probability distribution for observing an eigenvalue $m$ of $\hat{S}_z$, $P_m = \sum_{j=|m|}^{N/2}\mel{j,m}{\rho_{s}}{j,m}$, with $\ket{j,m}$ being the usual states of angular momentum $j$. We notice a bimodal distribution centered at two $m$ values, which agree with the stable mean-field solutions. This bimodal distribution is consistent with the existence of a first-order phase transition~\cite{binder_finite-size_1984}, here in a dissipative non-equilibrium setting. We estimate the phase-transition point from the middle points of the exact magnetization curves $\langle\hat{S}_z\rangle$ of increasingly large  values of $N$, finding it converges to $\Omega_{\mathrm{PT}}\approx 0.61\Omega_c$ (Appendix~\ref{app:PTpoint}).

Importantly, the first-order transition is seen in relevant physical observables beyond just the magnetization. For example, inter-atomic quantum correlations can be characterized  by the spin-squeezing parameter $\xi^2 = \min_\varphi \text{Var}[\hat{S}'_\varphi]N/|\langle \hat{\bm S} \rangle|^2$ \cite{ma_quantum_2011}
at steady state $\rho_s$. Here $\hat{S}'_\varphi$ is the spin-vector component directed at an angle $\varphi$ on the plane perpendicular to the mean spin $\langle\hat{\bm{S}}\rangle$ [with $\hat{\bm{S}}=(\hat{S}_x,\hat{S}_y,\hat{S}_z)$], and $\xi^2<1$ implies spin-squeezing entanglement useful in metrology. We observe in Fig. \ref{fig:res1}(d) that at the transition point the system jumps from an entangled spin-squeezed state to a non-squeezed state. Therefore, signatures of correlated physics, known for pure collective dissipation (Dicke case), exist before the transition $\Omega<\Omega_{\mathrm{PT}}$, where the magnetization $\langle\hat{S}_z\rangle$ coincides with the lower branch of the mean-field solution.

The above suggests that the first-order transition is in fact a transition between collective-spin and individual-spin physics, as can be characterized by large and small angular momenta representations, respectively. Indeed, we show this directly in Fig.~\ref{fig:res1}(c) by plotting the average of the total angular momentum operator $\hat{\bm S}^2 = \hat{S}_x^2+\hat{S}_y^2+\hat{S}_z^2, \quad \hat{\bm S}^2 \ket{j,m} = j(j+1)\ket{j,m}$. Before the transition, it is close to the maximal eigenvalue $j=N/2$ representing fully-symmetric, collective Dicke subspace, whereas at the transition it jumps close to zero for large enough $N$. The latter is consistent with a completely mixed state of $N$ independent spins~\cite{ostermann_breakdown_2023,agarwal_directional_2024}. Indeed, the results obtained with such a mixed state, as calculated in Appendix~\ref{app:mixed} are seen to exhibit excellent agreement with the exact solution at large enough $\Omega$ [Fig.~\ref{fig:res1}(c)].

These ideas are nicely manifested also in the properties of the collectively radiated light. We observe in Fig.~\ref{fig:res1}(e) that the average intensity $\langle \hat{S}_+ \hat{S}_-\rangle$ exhibits a sharp drop from a quadratic dependence on the drive amplitude, similar to that known from the collective Dicke case \cite{somech_heisenberg-langevin_2023,ferioli_non-equilibrium_2023}, to the constant value obtained for a mixed state of independent saturated atoms. Similarly, the photon correlations $g^{(2)}(0)=\langle \hat{S}_+ \hat{S}_+\hat{S}_-\hat{S}_-\rangle/\langle \hat{S}_+ \hat{S}_-\rangle^2$ jump from those consistent with coherent light ($g^{(2)}(0)=1$), as in the Dicke case \cite{somech_heisenberg-langevin_2023,somech_quantum_2024,sanchez_munoz_symmetries_2019}, to those of saturated independent atoms where $g^{(2)}(0)\rightarrow 2$.

%

\subsection{Conclusion: transition of competing dissipations}

The dissipative first-order transition discussed above is that between a steady-state phase dominated by correlated physics and a phase wherein the spins are largely independent. These two phases are associated, respectively, with purely collective and purely independent dissipations, thus mirroring the interplay between the two competing decay processes of Eq. (\ref{Eq:ME}).
We characterized both phases via all relevant physical observables of spins and radiation, including entanglement and correlations. In particular, the first-order transition from correlated to individual phase is marked by a sharp drop in the total angular momentum.
This is in stark contrast to the continuous, second order transition known in the driven Dicke case~\cite{drummond_volterra_1978,carmichael_analytical_1980,kessler_dissipative_2012,sanchez_munoz_symmetries_2019,barberena_driven-dissipative_2019}:
there, only purely-collective dissipation exists, so the total angular momentum is fixed to the maximal value $j=N/2$, and the individual spins are never fully mixed and independent, even at the saturated phase $\Omega/\Omega_c>1$.

\section{Quantum Bistability}
\label{sec:bistability}

\begin{figure}
    \centering
    \includegraphics[width=1\linewidth]{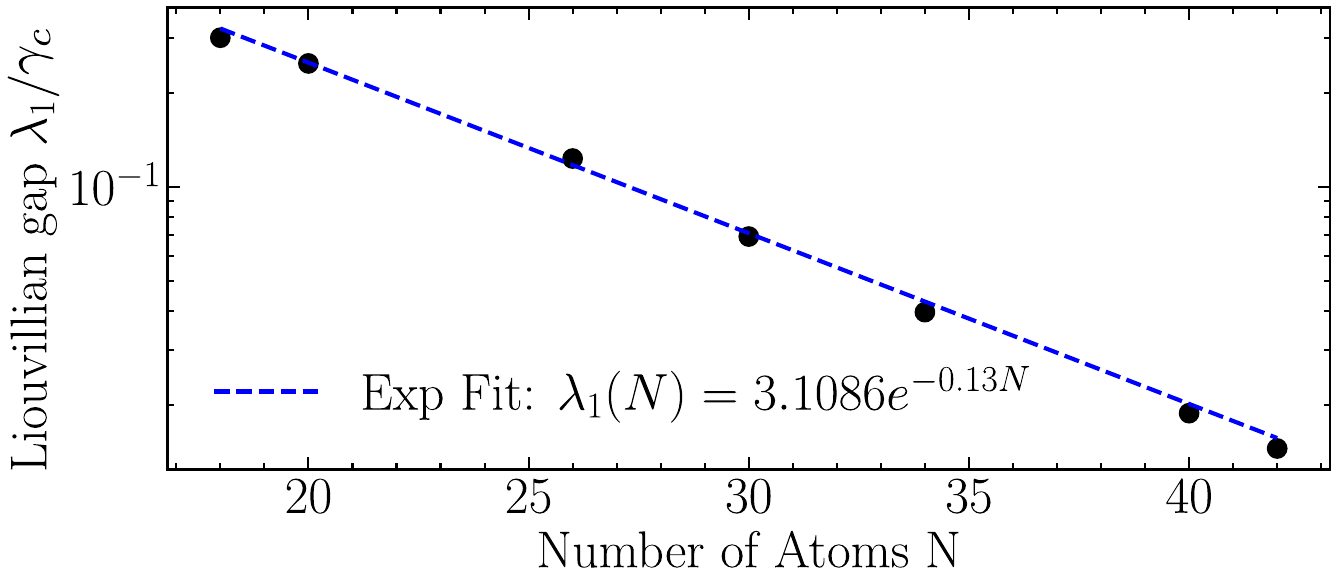}
    \caption{Liouvillian gap $\lambda_1$ as a function of the system size $N$ (semi-log scale). Here $\lambda_1$ is extracted from the long-time relaxation dynamics of $\langle \hat{S}_z(t) \rangle$ calculated exactly from Eq.~(\ref{Eq:ME}) at the critical point $\Omega = \Omega_{\mathrm{PT}}=0.61 \Omega_c$ (with $\gamma_c = 10 \gamma_s$, see text). The closing of the gap with $N$ is consistent with an exponential law.}
    \label{fig:lgap}
\end{figure}

We proceed to develop a deeper understanding of the first-order phase transition by directly finding the many-body states that correspond to the two phases. In particular, the mean-field analysis revealed a region of bistability between the phases (Sec.~\ref{sec:1spPT_mf}). We now study the meaning of this bistability in the quantum-mechanical sense, by relating the steady-state density matrix to two corresponding many-body states.

To this end, we characterize the dissipative phase transition by analyzing the eigenvalues $\lambda_i$ and eigenvectors $\rho_i$ of the Liouvillian superoperator ${\cal L}$ from Eq.~\eqref{Eq:ME}, following Ref.~\cite{minganti_spectral_2018}. The eigenvalue with the smallest non-vanishing real part is called the Liouvillian gap and denoted by $\lambda_{1}$. Recalling the uniqueness of the steady state  $\rho_{0}\equiv \rho_{s}$ ($\lambda_0=0$) at finite $N$, we can write the density matrix at long times $t\gg 1/\lambda_i$ $\forall i\neq 0,1$ as
\begin{equation}\label{eq:rho_time}
    \rho(t) = \rho_{s}+c_{1} \rho_{1} e^{-\lambda_{1} t},
\end{equation}
with the constant $c_{1}$ determined from initial conditions. The dissipative first-order phase transition then could be characterized by the closing of the Liouvillian gap, $\lambda_{1}\to 0$ for $N \to \infty$ at the critical drive $\Omega=\Omega_{\mathrm{PT}}$.
In this case, the kernel of the Liouvillian ${\cal L}$ becomes two-fold degenerate, leading to two stable phases.

We found the Liouvillian gap $\lambda_{1}$ at the critical point $\Omega_{\mathrm{PT}}\approx 0.61\Omega_c$ by fitting the long-time dynamics of the magnetization, calculated from Eq.~\eqref{Eq:ME} with $\gamma_c = 10 \gamma_s$, to $\langle\hat{S}_z(t)\rangle -\langle\hat{S}_z(\infty)\rangle\propto e^{-\lambda_{1}t}$ for different number of atoms $N$~\cite{vicentini_critical_2018}. The agreement of the resulting $\lambda_{1}$ values with those obtained by a direct diagonalization of  ${\cal L}$ was verified for $N\leq 18$ where the direct diagonalization is still feasible. The results are plotted in Fig.~\ref{fig:res1}(d), suggesting that the gap indeed closes exponentially with $N$, consistent with typical first-order transitions \cite{ptaszynski_dynamical_2024}.

\begin{figure}
    \centering
    \includegraphics[width=\columnwidth]{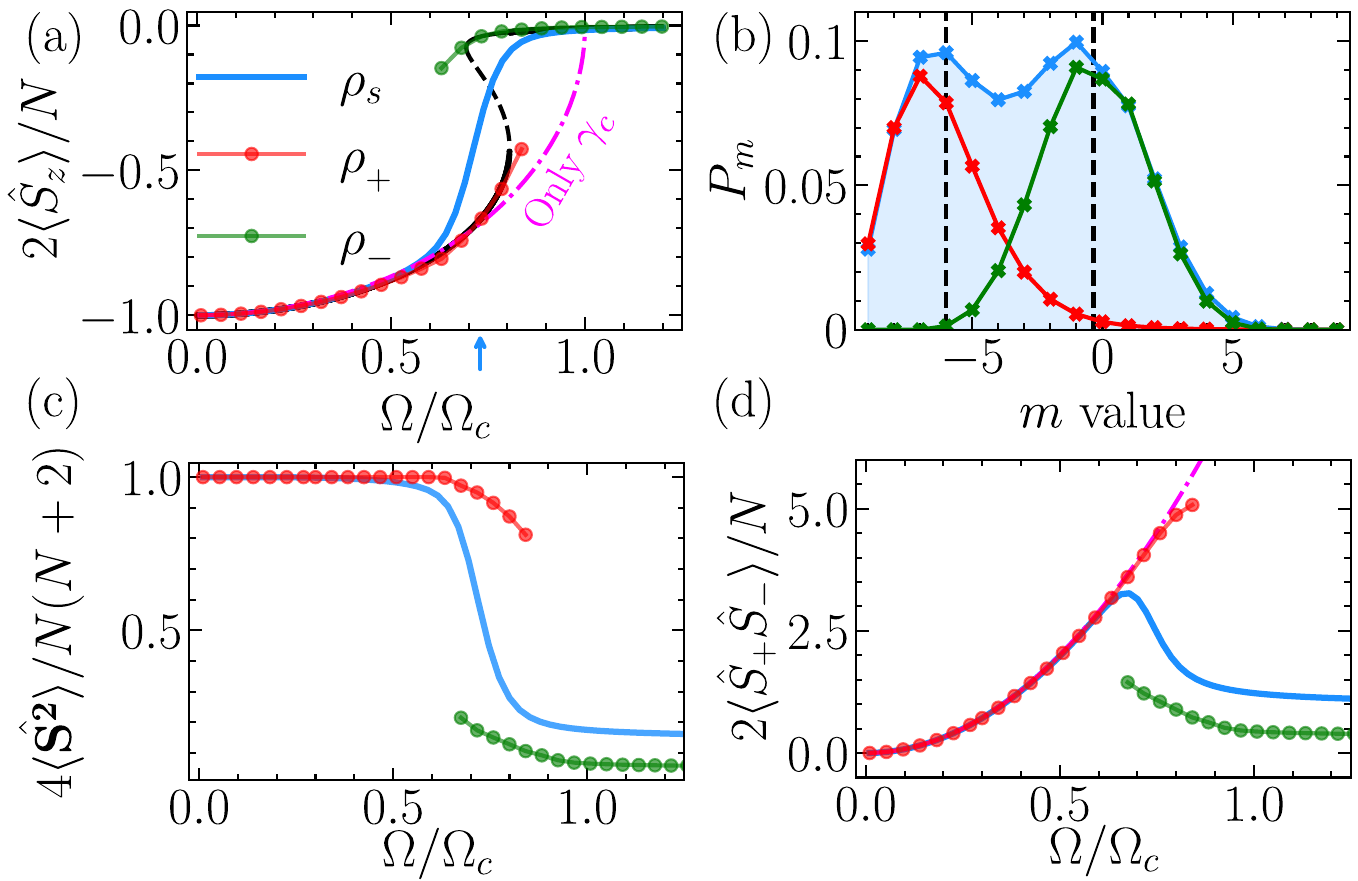}
    \caption{Quantum bistability: steady-state density matrix $\rho_s$ expressed as a mixture of two stable states $\rho_{\pm}$, as in Eq.~(\ref{eq:rho_ss_ppm}). The states $\rho_{\pm}$ are obtained numerically for $N = 18$ and $\gamma_c = 10\gamma_s$.
(a) Average magnetization calculated with $\rho_{+}$ ($\rho_{-}$) agrees with the lower- (upper-) branch mean-field solution.
(b) The magnetization distributions for $\rho_{\pm}$ are centered around their respective mean-field values, and their weighted sum (using $a_{\pm}$) approximately reproduces the bimodal distribution of $\rho_s$ (here for $\Omega = 0.73 \Omega_c$).
(c) Average total angular momentum $\langle\hat{\bm{S}}^2\rangle$ gives direct evidence that $\rho_+$ is associated with the collective phase (consistent with $j\approx N/2$) whereas $\rho_-$ with the individual-spin phase (consistent with low $j$'s). (d) Similar to (c), here for collective radiation intensity $\langle \hat{S}_+ \hat{S}_- \rangle$. The curve for $\rho_+$ exhibits excellent agreement with the result of pure collective dissipation (Dicke case $\gamma_s=0$, shown in magenta dashed-dotted line).
}
    \label{fig:results}
\end{figure}

The closing of the gap, $\lambda_1\rightarrow 0$, implies that $\rho_{1}$ tends to enter the kernel and that the steady state encodes two quantum states corresponding to two phases. To determine the two states we decompose $\rho_1$, which is seen from Eq. (\ref{eq:rho_time}) to be traceless, into two physical (trace-$1$) density matrices, $\rho_1=\rho_{+}-\rho_{-}$, with $\rho_+$ ( $\rho_-$) formed by the diagonal matrix containing the positive (negative) eigenvalues of $\rho_1$ ~\cite{minganti_spectral_2018}. The latter are orthogonal and approximately span the kernel, allowing to write
\begin{equation}\label{eq:rho_ss_ppm}
    \rho_{s}  \underset{N\to \infty}{\simeq}  a_+ \rho_{+} + a_- \rho_{-},\quad a_++a_- \simeq 1.
\end{equation}
Hence, the steady state is  given by a statistical mixture of two quantum many-body states. We demonstrate this for $N = 18$ by numerically finding $\rho_{1}$ as the eigenvector of the smallest nonzero eigenvalue of ${\cal L}$ and constructing $\rho_{\pm}$ from its diagonal form. In Fig~\ref{fig:results}(a), we plot the average of $\hat{S}_z$ taken with $\rho_{+}$ and $\rho_{-}$, showing their respective agreement with the lower and upper branches of the bistable mean-field solution. The correspondence to the two steady-state phases is further exhibited by the respective distributions of $m$ plotted for $a_\pm \rho_{\pm}$ in Fig.~\ref{fig:results}(b), whose sum recovers the bimodal distribution of $\rho_s$ to a very good approximation. Here $a_+ = 0.4$ and $a_- = 0.5$ were independently obtained from $a_{\pm}=\mathrm{tr}(\rho_s\rho_{\pm})/\mathrm{tr}[(\rho_{\pm})^2]$ noting their sum tends to $1$, as in (\ref{eq:rho_ss_ppm}), already for the moderate number of atoms $N=18$ reached here (see also Appendix~\ref{app:rho1}, Fig.~\ref{fig:apam_err}). 

The manifestation of the bistability as a mixture of two well-defined quantum states $\rho_\pm$ can be demonstrated for any physical observable. For example, in Fig.~\ref{fig:results}(c) we plot the mean of the total angular momentum $\hat{\bm{S}}^2$: its evaluation with $\rho_+$, $\text{Tr}[\hat{\bm{S}}^2\rho_+]$, is seen to correspond to a fully collective spin $j=N/2$ for drive stregths extending from $\Omega=0$ all the way to the right edge of the bistability region (beyond the phase transition point $\Omega_{\mathrm{PT}}$ found above with $\rho_s$). In contrast, for $\rho_-$ we observe values close to zero, consistent with a completely mixed state, which extend from the left edge of the bistabilty region to larger $\Omega$ (imperfections of the agreement with a total mixed state, seen for $N=18$, reduce with $N$).

Likewise, Fig.~\ref{fig:results}(d) shows the average intensity of collective radiation. Throughout the bistability region, the evaluation with  $\rho_+$ exhibits the quadratic dependence on $\Omega$ expected for the Dicke case, while the calculation with $\rho_-$ is consistent with the result of independent atoms. 
%

\subsection{Conclusion: transition between quantum states}

The quantum analysis of the bistability adds the following insights into the essence of the interplay between dissipations: Collective dissipation stabilizes the state $\rho_+$ while individual dissipation stabilizes $\rho_-$; accordingly, each state gives distinct predictions to the value of any physical observable, with $\rho_+$ showing signatures of correlated Dicke physics and $\rho_-$ resembling independent spins. Their mixture, $\rho_s=a_+\rho_+ + a_- \rho_-$, yields the average steady-state and exhibits a sharp transition between the states at a critical drive $\Omega_{\mathrm{PT}}$ (see also Appendix~\ref{app:rho1}).

The identification of the two phases with distinct quantum many-body states $\rho_{\pm}$ provides direct access to all relevant information on the phases and their transition.
In particular, by decomposing the states $\rho_{\pm}$ to their angular momentum components $|j,m\rangle$, we reveal that the two phases reside in different sections of the Hilbert space (Appendix~\ref{app:rho1}): $\rho_+$ is spanned entirely by states of large angular momentum close to $j=N/2$, implying approximate indistinguishablility between spins associated with collective physics. In contrast, $\rho_-$ is largely dominated by low $j$ values similar to the totally mixed state. This gives a direct evidence to the idea of a bistability of collective and individual physics. Notably, this perspective cannot be shown in principle from the mean-field analysis of the bistability, since states at the mean-field level are represented by a vector $(s_x,s_y,s_z)$ and are ambiguous at the full quantum level: e.g. the mean-field state $(0,0,0)$ can be associated with both $|j=N/2,m=0\rangle$ and $|j=0,m=0\rangle$.

\begin{figure}
    \centering
    \includegraphics[width=\columnwidth]{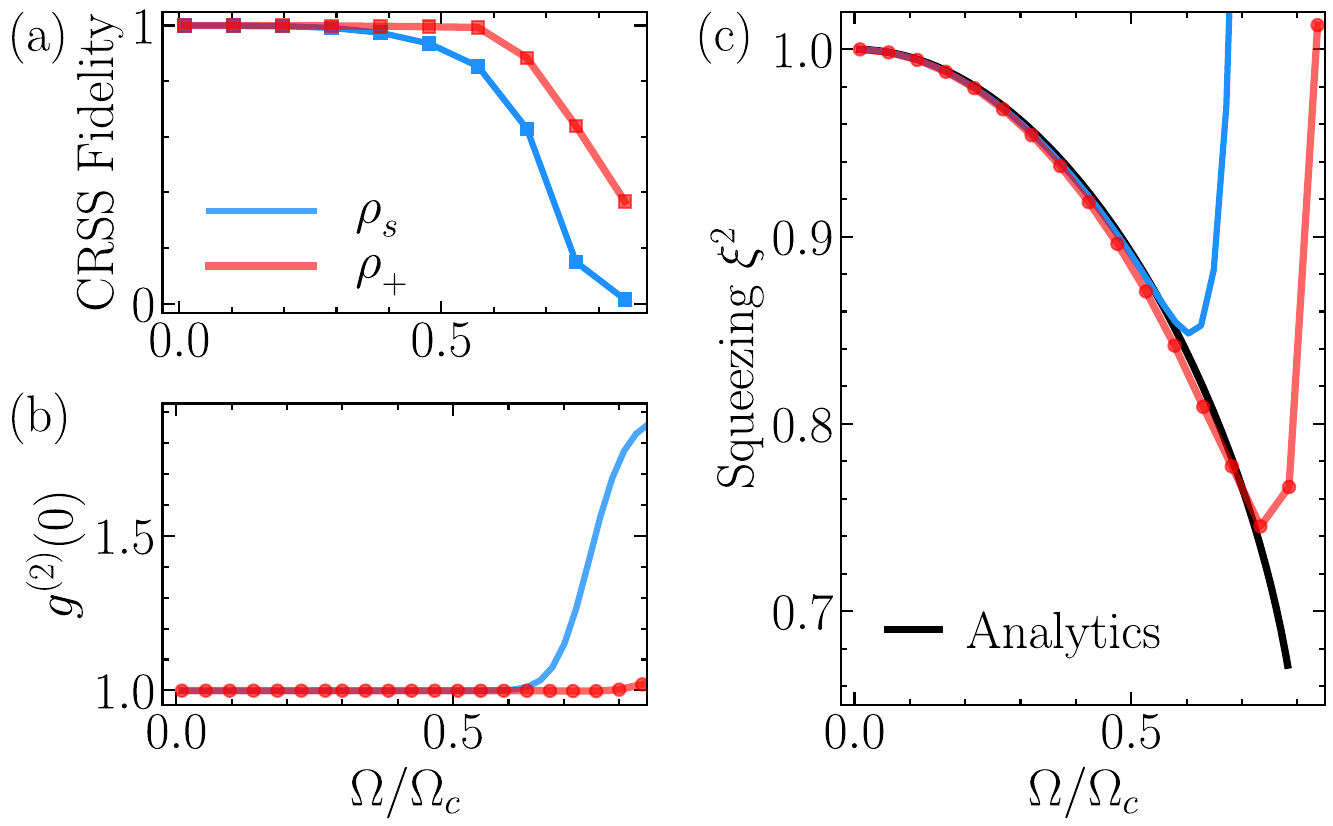}
    \caption{Emergence of Dicke-like correlated physics at the presence of decorrelating individual decay. (a) The fidelity of the steady-state component $\rho_+$ with the CRSS --- the entangled solution of purely collective dissipation ($\gamma_s=0$, Dicke)--- remains close to unity until the end of the bistability region (while that with the steady-state mixture $\rho_s$ drops already at the critical point).
    (b) Second-order correlations of collective radiation for $\rho_+$ are close to $1$, indicating coherent-state radiation as expected for CRSS. (c) Spin squeezing entanglement, $\xi^2<1$, is contained in the CRSS-like state $\rho_+$, in excellent agreement with the analytical result, Eq.~\eqref{eq:xi} (black line). Parameters taken in all plots are the same as those of Fig.~\ref{fig:results}}
    \label{fig:crss_squeezing}
\end{figure}

\section{Emergence of correlations: CRSS}
\label{sec:CRSS}
The quantum-bistability description, Eq. (\ref{eq:rho_ss_ppm}), allows to distinguish the component $\rho_+$ that shows signatures of collectivity. Therefore, the study of $\rho_+$ allows to address a main motivation of this work: namely, how correlated physics generated by collective dissipation emerges at the presence of individual decay.
To this end, we use CRSS theory for an analysis at the level of the quantum states themselves: we directly compare $\rho_+$ to the correlated CRSS that underlies pure collective dissipation.

First, we briefly review relevant aspects of CRSS \cite{somech_quantum_2024}.
Mathematically, a CRSS $|j,\alpha\rangle$ is given by the asymptotic eigenstate of the lowering operator of a large spin $j\gg 1$ with an eigenvalue $\alpha$, $\hat{S}_- |j,\alpha\rangle = \alpha |j,\alpha\rangle$ ($|\alpha|<j$). It turns out that the steady state of Eq. (1) for pure collective dissipation ($\gamma_s=0$) and drive strength $\Omega<\Omega_c$ is given by a CRSS of spin $j=N/2$ and eigenvalue $\alpha=(-i N/2)\Omega/\Omega_c$. Importantly, the spin properties of the CRSS are quantum, while its radiation properties are classical: it exhibits spin squeezing and pairwise entanglement between the $N$ constituent spins, while the scattered collective radiation $\hat{E}_c \propto \hat{S}_-$ is in a classical-like coherent state. CRSS is thus the many-body entangled spin state that encodes the correlated physics generated by pure collective dissipation (in contrast to the separable ``coherent spin state"~\cite{ma_quantum_2011}).

In Fig.~\ref{fig:crss_squeezing}(a) we directly compare $\rho_+$ to a corresponding CRSS by calculating the fidelity between the states. To this end, we choose $\Omega$ inside the bistabiity region, find $\rho_s$ and $\rho_+$, and construct the corresponding CRSS of eigenvalue $\alpha=(-i N/2)\Omega/\Omega_c$ using the CRSS expansion formula in $|j=N/2,m\rangle$ states \cite{somech_quantum_2024}. We observe very high fidelities, dropping only towards the end of the bistability region. Remarkably, this suggests that all quantum properties of collective CRSS physics should be exhibited by the stable state $\rho_+$, including its coherent radiation and spin-squeezing entanglement. Indeed, in Fig.~\ref{fig:crss_squeezing}(b) we plot the photon correlations evaluated with $\rho_+$, observing excellent agreement with the coherent-state result $g^{(2)}(0)=1$ in the corresponding region of bistability. We also observe in Fig.~\ref{fig:crss_squeezing}(c) the existence of spin squeezing ($\xi^2<1$), which improves with $\Omega$ until its degradation near the edge of the bistability region.

\subsection{Optimal spin squeezing}
The dependence of the spin squeezing $\xi^2$ on the drive $\Omega$ seen in Fig.~\ref{fig:crss_squeezing}(c) shows that $\xi^2$ reaches an optimal (minimal) point at $\Omega$ close to the edge of the bistability region. More insight into the optimally achievable squeezing in $\rho_+$ can be gained by estimating the spin squeezing analytically using a Heisenberg-picture approach~\cite{somech_heisenberg-langevin_2023}.
To this end, we begin with the Heisenberg-Langevin equations for $\hat{\sigma}_n$ and $\hat{\sigma}_n^z$ which correspond to Eq.~\eqref{Eq:ME}.
 Linearizing the equations for small fluctuations around the lower branch of the mean-field solution (\ref{eq:gammaneq0}), which corresponds to $\rho_+$, and using the Holstein-Primakoff transformation, we find the dynamical equation of collective-spin fluctuations (Appendix~\ref{app:spin_squeezing}),
\begin{align}\label{eq:a_dot}
	&\dot{\hat{a}} =\qty(\frac{\Gamma j_{\rm MF} \cos \theta}{N}  -{\gamma\over 2}) \hat{a}-{\gamma\over 4}\sin^2\theta (\hat{a} -\hat{a}^\dagger)+\sqrt{j_{\rm MF}\over 2N}\times \notag \\& \hspace{-0.1 cm}\qty[\cos\theta(\hat{F}+\hat{F}^\dagger+\hat{\eta}+\hat{\eta}^\dagger)+(\hat{F}^\dagger-\hat{F}+\hat{\eta}^\dagger-\hat{\eta})].
\end{align}
Here $\hat{a} = \frac{1}{\sqrt{N}}\sum_n \hat{a}_n$, $\hat{F} = \sqrt{N}\hat{f}$ and $\hat{\eta} = \frac{1}{\sqrt{N}}\sum_n \hat{\eta}_n$, where $\hat{a}_n$ is the bosonic fluctuation of the spin of atom $n$, and $\hat{f}$ ($\hat{\eta}_n$) is the vacuum noise associated with collective (individual) decay $\gamma_c$ ($\gamma_s$). The effective spin size $j_{\rm MF} = |\bm{s}|N/2$ and excitation angle $\theta=\arccos(s_z/|\bm{s}|)$ with $|\bm{s}|=\sqrt{s_x^2+s_y^2+s_z^2}$ are given by the lower-branch mean-field solution $s_{x,y,z}$. Within this formulation, the spin squeezing is obtained as
\begin{equation}\label{eq:xi}
	\xi^2 = 1+2\qty(\expval{\hat{a}^\dagger \hat{a}}-\abs{\expval{\hat{a}^2}}).
\end{equation}
Solving Eq.~\eqref{eq:a_dot} for $\hat{a}$ in steady state, and using the correlations of vacuum noises $\langle\hat{f}(t)\hat{f}^{\dag}(t')\rangle=\gamma_c\delta(t-t')$, $\langle\hat{\eta}_n(t)\hat{\eta}_m^{\dag}(t')\rangle=\gamma_{s}\delta_{nm}\delta(t-t')$, we obtain an analytical expression for $\xi^2$ in terms of the mean-field parameters $\Gamma$ and $\gamma$ and the lower-branch solution $s_{x,y,z}$. The full analytical expression for $\xi^2$ is given in Appendix~\ref{app:spin_squeezing} and plotted in Fig.~\ref{fig:crss_squeezing}(c): we observe excellent agreement with the squeezing calculated for the stable state $\rho_{+}$, extending towards the end of the bistability region.
In addition, we note the agreement with the results of the full steady state $\rho_s$ before the bistability region [see also Fig.~\ref{fig:res1}(c)].

\begin{figure*}[ht!]
    \centering
    \includegraphics[width = \linewidth]{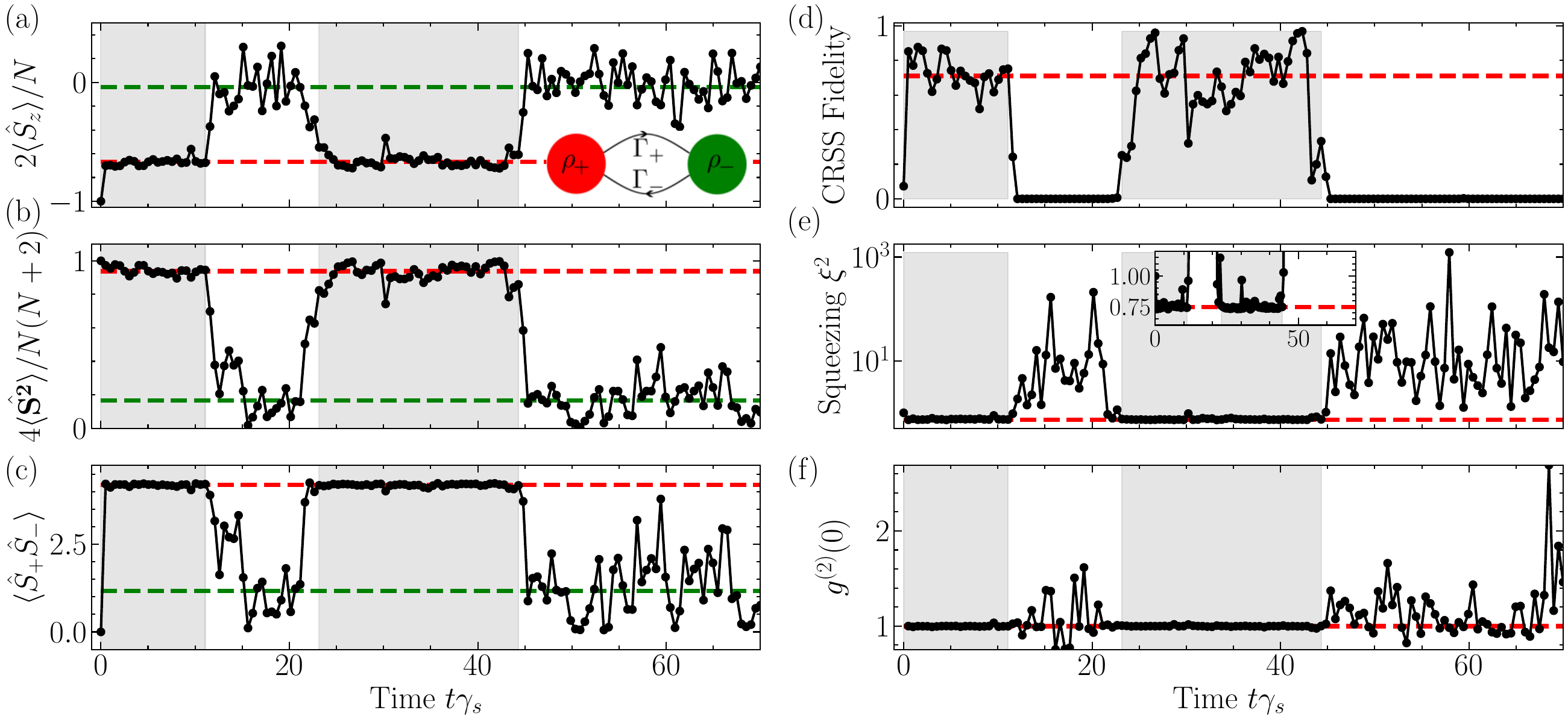}
       \caption{Quantum switching between stable states $\rho_{\pm}$ manifested in quantum-trajectory simulations, performed using QuTip~\cite{johansson_qutip_2013} for $N = 18$, $\gamma_c = 10\gamma_s$, and $\Omega = 0.73\Omega_c$, with all atoms initially in the ground state. All plots are for the same single trajectory wavefunction, so that switching between values corresponding to $\rho_{\pm}$ are seen simultaneously in the various observables:
(a,b,c) Averages of the magnetization $\hat{S}_z$, total angular momentum $\hat{\bm S}^2$ and collective radiation intensity $\hat{S}_+ \hat{S}_- $ evaluated on the trajectory wavefunction, compared to their calculation with $\rho_+$ and $\rho_-$ (red and green dashed lines, respectively). (d) CRSS fidelity of the trajectory wavefunction compared to that of $\rho_+$ (red dashed curve). High fidelity with CRSS is seen for the corresponding $\rho_+$-matched intervals identified in (a,b,c).
(e) Spin squeezing along the same trajectory compared to the result with $\rho_+$ (red dashed curve). Inset: Zoomed-in view of the region $0.5 < \xi^2 < 1.25$. Squeezing correlations $\xi^2<1$ exist in the CRSS-like intervals from (d). (f) Same as (e) for $g^{(2)}(0)$ correlations of collective radiation.}
       \label{fig:quant_jump}
\end{figure*}

For studying the spin squeezing at large particle number $N$, we consider the limiting case $\Gamma \gg \gamma$: substituting the corresponding mean-field values from Eq.~\eqref{eq:sz_an}, we get
\begin{equation}\label{eq:squez_G_large}
   \hspace{-0.25 cm} \xi^2 = \frac{1+\sqrt{1-2\Omega^2/\Omega_c^2}}{\sqrt{2}\sqrt{1+\Omega^2/\Omega_c^2+\sqrt{1-2\Omega^2/\Omega_c^2}}}\xrightarrow[\Omega = \frac{\Omega_c}{\sqrt{2}}]{} \frac{1}{\sqrt{3}}.
\end{equation}
The optimum is obtained at the edge of the bistability region at $\Omega=\Omega_c/\sqrt{2}$ yielding a best achievable squeezing of $1/\sqrt{3}$. Notably, this optimum does not depend on $N$, in contrast to the scaling $\xi^2\sim N^{-1/3}$ obtained for pure collective dissipation using CRSS theory \cite{somech_quantum_2024}. This is since the squeezing of CRSS improves as we approach $\Omega=\Omega_c$, whereas $\rho_+$, although equivalent to a CRSS to an excellent approximation, becomes irrelevant much before, at $\Omega=\Omega_c/\sqrt{2}<\Omega_c$, thus setting the best achievable squeezing to a constant $1/\sqrt{3}$.

Importantly, the above analysis of correlated-physics properties is performed for $\rho_+$, whereas the full steady state is $\rho_s$. In the following we discuss the situations at which the physics captured by $\rho_+$ is indeed the observable physics, thus making the above analysis the relevant one.

\section{Quantum switching dynamics}
\label{sec:SS_dyn}
While the total steady-state $\rho_s$ encodes the two stable states $\rho_{\pm}$ only by their statistical mixture (\ref{eq:rho_ss_ppm}), the quantum bistability has an observable meaning directly revealed in dynamics. Recalling that the long-time dynamics, Eq.~\eqref{eq:rho_time}, is spanned by $\rho_{s}$ and $\rho_{1}$ and hence approximately by $\rho_{\pm}$, we use $\rho_1=\rho_+-\rho_-$ and~\eqref{eq:rho_ss_ppm} in Eq.~\eqref{eq:rho_time}, obtaining the density matrix at long-times,
\begin{equation}\label{eq:rho_t_pm}
    \rho(t) = p_+(t) \rho_{+}+p_-(t) \rho_{-},
    \quad p_\pm(t) = a_\pm \pm c_{1} e^{-\lambda_{1}t}.
\end{equation}

The probabilities $p_{\pm}(t)$ to find the system in states $\rho_{\pm}$ then dynamically evolve as
\begin{equation}\label{eq:p_pm}
    \dot{p}_\pm(t) = -\Gamma_{\pm} p_\pm(t) + \Gamma_{\mp} p_\mp(t),
    \quad \Gamma_{\pm} = \lambda_{1}a_\mp.
\end{equation}
Notably, Eq.~\eqref{eq:p_pm} describes a two-state Markov process of stochastic jumps between these states at rates $\Gamma_{\pm}$ [inset of Fig.~\ref{fig:quant_jump}(a)]. However, unlike a classical Markov process, the jumps are activated by quantum-vacuum noise and occur between two states in Liouville space, each representing a many-body state with its own quantum statistics and properties. Therefore, we predict a physical reality of dynamical switching not only of a specific order parameter, as studied in previous works~\cite{lee_collective_2012,wilson_collective_2016,gelhausen_dissipative_2018,brookes_critical_2021,minganti_dissipative_2023,gabor_quantum_2023}, but of the full quantum properties including correlations and entanglement.

This prediction is nicely manifest in the simulation of single trajectories from the quantum stochastic unraveling of Eq.~\eqref{Eq:ME}~\cite{molmer_monte_1993}. In Fig.~\ref{fig:quant_jump}, we observe the jumps between the two values corresponding to $\rho_{\pm}$ for various observables of atoms and light: magnetization $\langle \hat{S}_z \rangle$, total angular momentum $\langle \hat{\bm S^2} \rangle$,  radiation intensity $\langle \hat{S}_+ \hat{S}_-\rangle$, fidelity with CRSS, spin squeezing, and photon correlations $g^{(2)}(0)$. The fluctuations around the two values, originating in the statistics of the corresponding states $\rho_{\pm}$, decrease with the system size $N$.

The direct observation of quantities such as $\hat{S}_z$ and $\hat{S}_+ \hat{S}_-$ along quantum trajectories is experimentally accessible through photon detection at the corresponding ports of individual and collective dissipation, as shown in Fig.~\ref{fig:sketch}(a). The resulting photocount can be simulated by tracking the times at which individual or collective jump operators act (Appendix~\ref{app:photon_counting}). Using this approach, we observe that the switching between the states $\rho_\pm$ typically involves a burst of many individual jump events occurring within a short time window (Appendix~\ref{app:photon_counting}, Fig.~\ref{fig:short_jump}).
The need for many individual jumps is nicely understood recalling that $\rho_\pm$ differ dramatically in their span of total angular momentum $j$. Since collective jumps cannot change $j$, the switching must involve individual jumps; and since the difference in $j$ is macroscopic, this requires many such jumps. The switching between $\rho_\pm$ is hence performed by an effective \emph{many-body} quantum jump consisting of multiple individual jumps of different atoms~\cite{lee_collective_2012}. This is in contrast to the switching between polariton states in photon blockade systems, where a single jump operator may be sufficient~\cite{carmichael_breakdown_2015}.

Importantly, the switching rates $\Gamma_{\pm}\propto \lambda_1$ tend to zero at the critical point as $\lambda_1\rightarrow 0$ [Fig.~\ref{fig:lgap}], implying that the system resides for exceedingly long times in either of the states $\rho_{+}$ or $\rho_{-}$. For the former, recalling the resemblance of $\rho_{+}$ to a CRSS [Figs.~\ref{fig:crss_squeezing} and~\ref{fig:quant_jump}(d-f)], this leads to the following remarkable conclusion: namely, that the practically observable physical reality is similar to that of the Dicke problem of purely correlated decay, exhibiting metrologically useful spin squeezing entanglement, even at the presence of decorrelating individual decay. While it might be expected that Dicke physics emerges at short times $t \ll\gamma_s^{-1}$ before individual decay is noticeable, here we surprisingly find that this can be the case also at the true steady state for $t\gg \gamma_s^{-1}$. This is further discussed in the following section.

We stress that Eq. (\ref{eq:p_pm}), which entails the switching dynamics, was derived by us directly from the theory of dissipative phase transitions, and is used here with our knowledge of $\rho_{\pm}$ to \emph{predict} the simulated trajectories of various quantum properties seen in Fig.~\ref{fig:quant_jump}. This is in contrast to previous works where a similar equation was presented as an effective model justified \emph{posteriori} by simulated trajectories of a specific order parameter or its bimodal distribution~\cite{savage_single_1988,wilson_collective_2016,brookes_critical_2021}.
%

\section{Discussion: observable physics}
\label{sec:exp}
We turn to discuss important consequences on experiments: the observability of collective Dicke physics and of the bistability, including the hysteresis-based preparation of the correlated $\rho_+$ as a quasi-stationary state.

\subsection{Dicke physics}
While the Dicke model describes purely collective dissipation, its associated physics should be observable in realistic systems under some conditions. We now use our analysis to address this. From Eq.~\eqref{Eq:ME}, one intuitively identifies the timescale $(N\gamma_c)^{-1}$ of purely collective relaxation and the time $\gamma_s^{-1}$ when decorrelating individual decay becomes significant. When the separation of timescales $N \gamma_c\gg \gamma_s$ exists, one then expects to be able to observe collective physics approximately governed by Eq.~\eqref{Eq:ME} with $\gamma_s=0$ (driven Dicke model). A well-known example is the transient superradiant burst of freely decaying atoms, as measured in various systems~\cite{norcia_superradiance_2016,ferioli_laser-driven_2021}, and predicted beyond the purely symmetric case when there exists a dominant collective mode~\cite{gross_superradiance_1982,lin_chapter_2012,masson_many-body_2020,masson_universality_2022,mok_universal_2025}. Here, however, we are interested in robust steady-state phases rather then in transient effects; the relevant question is then, under which conditions one can observe the correlated CRSS --- the steady state of the driven Dicke model ($\gamma_s= 0$) --- at the presence of decorrelating individual decay ($\gamma_s\neq 0$).

\begin{figure}
    \centering
    \includegraphics[width=\columnwidth]{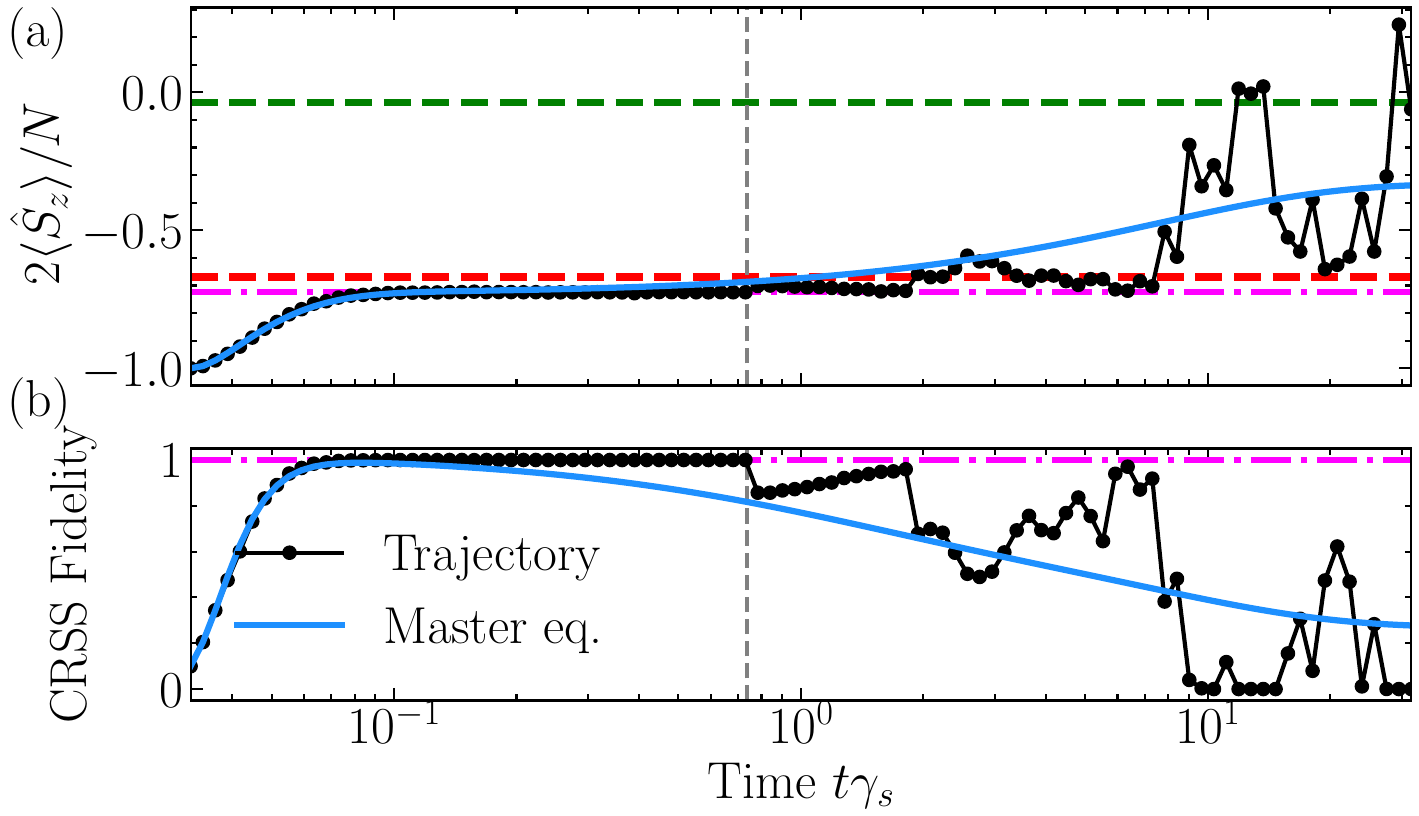}
    \caption{Transient dynamics of the time-dependent density matrix solution of Eq.~\eqref{Eq:ME} (blue solid curve) and of a specific quantum trajectory, for: (a) magnetization $\langle \hat{S}_z \rangle$, and (b) fidelity with CRSS (parameters: $N = 18$, $\Omega/\Omega_c = 0.73$, $\gamma_c=10 \gamma_s$). Fast relaxation to the Dicke solution, observed at $t\gamma_s\sim 10^{-2}$, is followed by a relaxation to the true steady state at a time-scale $t\gamma_s\sim 10$. The onset of the departure from Dicke physics and the approach to steady state is identified by the first individual quantum jump marked at $t\gamma_s\approx0.75$. This is where the CRSS fidelity first drops from unity.}
    \label{fig:short_time}
\end{figure}

We discuss two answers to this question. First, one intuitively expects to observe Dicke-like physics for an experiment duration $T$ that is short enough to ignore individual decay, $T\ll \gamma_s^{-1}$, yet long enough to reach the CRSS steady state, $T\gg (N\gamma_c)^{-1}$. We examine how this works in Fig.~\ref{fig:short_time} by plotting the exact transient dynamics of both the magnetization and the CRSS fidelity: While the true steady state, discussed in detail, is reached at a long times $t \gamma_s \gtrsim 10$, we see that at very short times $t\gamma_s\sim (N\gamma_c/\gamma_s)^{-1}\sim 10^{-2}$ the system reaches the CRSS, and remains there until the onset of relaxation to steady state at $t\gamma_s \sim 1$. Further insight into the departure from purely collective Dicke physics is given by the simulated quantum trajectory shown in Fig. ~\ref{fig:short_time}: after a very quick relaxation the state \emph{identically} becomes a CRSS (fidelity $=1$), and it remains so for quite a while. This is since the CRSS is an eigenstate of the collective dissipation $\hat{S}_-$. The departure from pure Dicke physics then occurs only after the first individual jumps, marked at $t \gamma_s \approx 0.75$ in Fig.~\ref{fig:short_time}. At this time, the CRSS fidelity immediately drops below $1$, heralding the onset of the relaxation to steady state. These results provide a direct evidence that $\gamma_s^{-1}$ is indeed the timescale that determines the system's departure from pure Dicke physics. Interestingly, this timescale is only revealed by the full quantum dynamics and does not appear in the mean-field approximation. This is since the mean-field assumption of no correlations, which is wrong at initial times, introduces the individual timescale  $\gamma=\gamma_s+\gamma_c$ instead of  $\gamma_s$ (see above and Appendix~\ref{app:MF}), hence fails to predict the true transient dynamics.

Beyond the short-time regime, we recall that correlated Dicke-like physics can in fact emerge even at the true steady state at long times $T\gg \gamma_s^{-1}$. Considering that at these times the decorrelating individual decay is significant, this presents a much less intuitive result, which nevertheless becomes clear from our analysis. Namely, since the steady state contains the CRSS-like state $\rho_+$, we predict the observation of CRSS Dicke-like physics either via quantum switching or as a quasi-stationary state in a macroscopic systems (where the switching rate vanishes). This entails an intriguing consequence on the interpretation of experiments: the observation of Dicke-like physics does not always imply negligible individual decay, but may also imply that the system retains correlations at the presence of individual decay, as per $\rho_+$.

These ideas are relevant for current experiments both in cavities~\cite{norcia_superradiance_2016,SK} and possibly in free space~\cite{ferioli_non-equilibrium_2023}. In particular, a recent experiment~\cite{song_dissipation_2024} probed mean-field observables both at short and long times. In this respect, our predictions provide valuable information on the full quantum state and correlations beyond mean values.

\subsection{Bistability and hysteresis}

Beyond the signature of the bistability as a bimodal distribution [Fig.~\ref{fig:res1} (b)], we discussed its direct observation via quantum switching dynamics between the states $\rho_\pm$ (Sec.~\ref{sec:SS_dyn}). We recall that the switching rate is proportional to the Liouvillian gap $\lambda_1$, which diminishes exponentially with the atom number $N$. This makes the observation of the switching relevant for mesoscopic systems, such as atom arrays in cavities~\cite{SK}, comprised of up to dozens of atoms or so.

Considering experiments with a macroscopic number of atoms, such as those from Refs.~\cite{norcia_superradiance_2016,rivero_quantum_2023,song_dissipation_2024} with $N \sim 10^5$, the switching rate practically vanishes. We then expect that the system settles in either of the states $\rho_+$ or $\rho_-$, the identity of which is determined by the initial conditions of the experiment. For a given $\Omega$ within the bistabilty region, the preparation of either $\rho_+$ or $\rho_-$ for different initial conditions forms a direct evidence of the bistability; when repeated for all values of $\Omega$ this traces a hysteresis loop.

\begin{figure}
    \centering
    \includegraphics[width=\columnwidth]{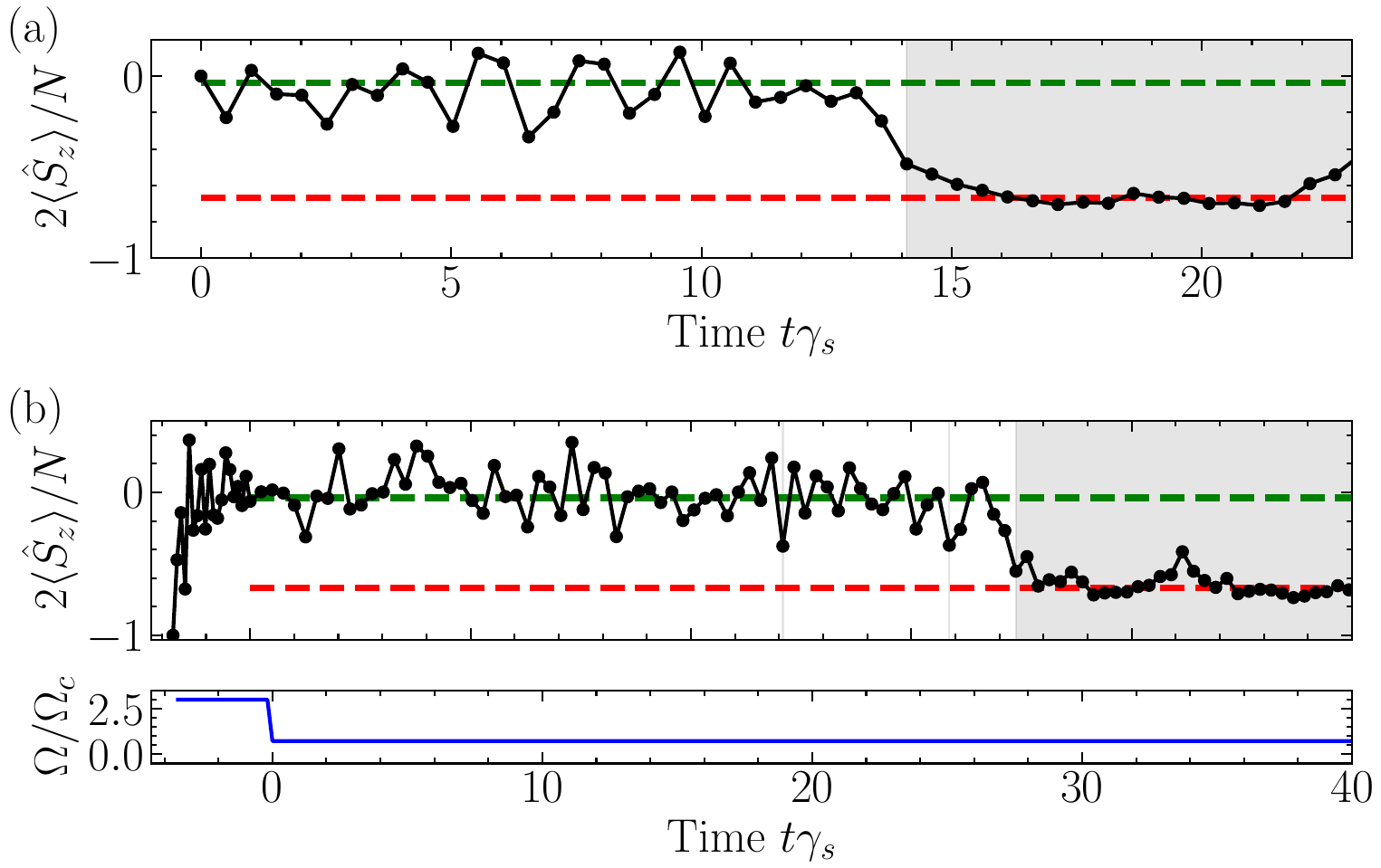}
    \caption{Quantum trajectories demonstrating the preparation of the state $\rho_-$ (for $N = 18$, $\gamma_{c} = 10\gamma_s$, and $\Omega = 0.73\Omega_c$). Indication of $\rho_-$ is given by the corresponding magnetization value (green dashed line). For macroscopic systems, $N\gg 1$, the switching, after the initial relaxation to $\rho_-$, is inhibited. (a) Preparation from an initial singlet-product state (text). (b) Preparation using a quench protocol. Starting with all atoms in the ground state, the driving field is changed as shown in the lower panel: first, it is set to $\Omega = 3\Omega_c$, and after a short relaxation it is changed abruptly to $\Omega = 0.73 \Omega_c$ causing the system to settle in $\rho_-$ as in (a).}
    \label{fig:quench}
\end{figure}

The question is then how to determine the initial conditions that yield $\rho_+$ or $\rho_-$ as a quasi-stationary state; this could be useful for the preparation of the entangled state $\rho_+$ in favor of the uncorrelated $\rho_-$. To this end, we note: (i) the short-time dynamics are governed by collective dissipation $\hat{S}_-$ that does not change the total angular momentum $j$; (ii) the onset of steady state is reached when the first individual jumps occur and change $j$. Given an initial state, the system will then first relax to either of the states $\rho_+$ or $\rho_-$ whose total angular momentum is closest to that of the initial state, since this requires fewer individual jumps. Now, considering that $\rho_+$ is close to a CRSS, its angular momentum is approximately $j=N/2$, whereas since $\rho_-$ is close to a total mixed state, its average angular momentum is very small, as seen in Fig.~\ref{fig:results}(c).

We now show how these principles work. Considering first the preparation of $\rho_+$, one may e.g. begin with a state of total angular momentum $j=N/2$, close to that of $\rho_+$. Indeed, initializing the system at the ground state of all atoms, $|j=N/2,m=-N/2\rangle$, we observe in Fig.~\ref{fig:quant_jump} and Fig.~\ref{fig:short_time} that after the onset of steady state at $t\gamma_s > 1$, the magnetization and CRSS fidelity are consistent with those of $\rho_+$. For macroscopic $N$, the subsequent switching to the state $\rho_-$ seen in Fig.~\ref{fig:short_time} does not occur and the system effectively stays in $\rho_+$. We obtain a similar result for the initial state $|j=N/2,m=0\rangle$ (not shown), highlighting that it is the total angular momentum $j$ that matters and not its components such as $\hat{S}_z$.

Likewise, for the preparation of $\rho_-$, consider first the initial state with $j=0$ given by the product of singlets, $|\psi_s\rangle\propto(\hat{\sigma}_1^\dagger-\hat{\sigma}_2^\dagger)(\hat{\sigma}_3^\dagger-\hat{\sigma}_4^\dagger)\dots|gg...g\rangle$.
A quantum trajectory for this initial state is shown in Fig.~\ref{fig:quench}(a), observing that the system indeed first relaxes to $\rho_-$.
A preparation method that does not require the initial entangled state $|\psi_s\rangle$, is provided by the quench protocol shown in Fig.~\ref{fig:quench}(b). First, one sets the drive $\Omega$ well above $\Omega_c$, for which the steady-state is a separable total mixed state of saturated atoms whose average total angular momentum is small [Fig.~\ref{fig:res1}(c), Appendix~\ref{app:mixed}]. After the relaxation to this small angular-momentum state, one then abruptly tunes the drive $\Omega$ to a desired value within the bistability region and the system relaxes to $\rho_-$, as seen in Fig.~\ref{fig:quench}(b).

\section{Conclusions}
We addressed the problem of collective dissipation at the presence of local dissipation, in steady state. We found that a correlated CRSS-like many-body state emerges as a component of a quantum bistability, discussed its associated entanglement, and the observability of this physics in different regimes.

These results provide a new perspective on the physics of collective dissipation beyond the Dicke indistinguishablity symmetry, by revealing the mechanism for the emergence of Dicke-like correlations at the presence of decorrelating local decay. As such, this work establishes an essential first step in a systematic study of the role and potential of collective dissipation phenomena in realistic  systems. In particular, considering quantum-technological platforms realized with ensembles or arrays of atom-like emitters in cavities or free space~\cite{kaluzny_observation_1983,bloch_ultracold_2005,barredo_atom-by-atom_2016,endres_atom-by-atom_2016,angerer_superradiant_2018,norcia_superradiance_2016,ferioli_non-equilibrium_2023,rivero_quantum_2023,kersten_self-induced_2024}, the identification of distinct effective channels of collective and individual decay will allow to map complex problems to the model studied here. 

Our analysis relied on CRSS theory for characterizing Dicke-like physics by associating it with a well-defined quantum state whose description in terms of angular momentum states is available~\cite{somech_quantum_2024}. This allowed to tackle properties well-beyond mean field both in terms of fidelities and correlations. Apart from CRSS itself, we note that our exact quantum analysis was essential to reveal several predictions that are unattainable using mean field. These do not only include correlation, entanglement and fluctuation properties which are clearly not accounted for by mean-field, but also certain crucial dynamical properties: e.g. both the time-scale $\gamma_s^{-1}$ of the onset of individual decay, which is absent from the mean-field dynamics, and also the prediction of the quasi-stationary state the system reaches, which must include information on the total angular momentum subspace.

\begin{acknowledgments}
We acknowledge fruitful discussions with Daniel Goncalves-Romeu and Darrick Chang, and financial support from the Israel Science Foundation (ISF), the Directorate for Defense Research and Development (DDR\&D), the Minerva
Foundation with funding from the Federal German Ministry for Education and Research, the Center for New Scientists at the Weizmann Institute of Science, and the Council for Higher Education (Israel). This research is made possible in part by the historic generosity of the Harold Perlman Family.

\end{acknowledgments}

\appendix

\section{Quantum optical derivation of model}
\label{app:HL_eqs}
We present a derivation of the model corresponding to Eq.~\eqref{Eq:ME}. For concreteness, we consider the cavity realization from Fig. 1(a), and employ a Heisenberg-Langevin approach which is completely equivalent to the master equation~\eqref{Eq:ME} but that is better suited for later analytical calculations. We consider $N$ two-level atoms inside a driven one-sided optical cavity. We assume that atoms are identically coupled to the cavity mode (e.g. located at cavity antinodes). The Hamiltonian for the atoms and the cavity mode is
\begin{align}
    \label{Seq:H_sys}
    {\cal H}_S &= \hbar \omega_0 \sum_{n}\hat{\sigma}_n^\dagger \hat{\sigma}_n  + \hbar \sum_n (g^*\hat{c}^\dagger \hat{\sigma}_n +g{\hat \sigma}_n^\dagger \hat{c}) \nonumber\\
    &\quad + \hbar \omega_c \hat{c}^\dagger \hat{c}+ \hbar  (\Omega_L \hat{c}^\dagger e^{-i\omega_L t}+\Omega_L^* \hat{c} e^{i\omega_L t}),
\end{align}
where $\hat{\sigma}_n$ is the Pauli lowering operator for atom $n$ with resonant frequency $\omega_0$, $\hat{c}$ is the boson lowering operator for the cavity mode of frequency $\omega_c$ and $g$ is a dipole coupling strength identical for all atoms. The external laser comes from the left side of the cavity with amplitude $\Omega_L$ and frequency $\omega_L$.

The cavity mode is coupled through the left mirror to the 1D continuum of the propagating photon modes with the wave numbers $k$ and frequencies $\omega_k = c \abs{k}$ ($c$ being the speed of light)
\begin{equation}\label{Seq:H_cav}
    {\cal H}_R = \hbar \sum_{k<0} \omega_k \hat{b}_k^\dagger \hat{b}_k, \quad {\cal H}_{SR} = \hbar \sum_{k<0}(g_c \hat{b}_k^\dagger \hat{c}+g_c^* \hat{c}^\dagger \hat{b}_k),
\end{equation}
where $\hat{b}_k$ are the corresponding boson lowering operators and $g_c$ the coupling strength.

Emission to off-axis modes outside the cavity is accounted for by approximating these modes as free-space modes described by the lowering operators $\hat{a}_{\bm k,\mu}$ and frequencies $\omega_{\bm k} = c \abs{\bm k}$, with ${\bm k}$ the wavevector and $\mu$ the polarization:
\begin{multline}
    \label{Seq:H_F}
    {\cal H}_F = \hbar \sum_{\bm k,\mu}  \omega_{\bm k} \hat{a}_{\bm k,\mu}^\dagger \hat{a}_{\bm k,\mu},\\ {\cal H}_{SF} = \hbar \sum_{\bm k,\mu}\sum_n \qty(g^*_{\bm k,\mu,n}\hat{a}_{\bm k,\mu}^\dagger\hat{\sigma}_n +g_{\bm k,\mu, n}  \hat{\sigma}_n^\dagger \hat{a}_{\bm k,\mu}),
\end{multline}
where $g_{\bm k,\mu,n} \propto e^{i \bm k \cdot \bm r_n}$ is the dipole coupling whose dependence on the atomic position $\bm r_n$ follows from the free-space plane waves $\bm k$.

Following the standard procedure, we eliminate the reservoir modes to obtain the equations on the atomic variables. We formally solve the Heisenberg equation on $\hat{b}_k$ and insert the solution into the Heisenberg equation for the cavity mode $\hat{c}(t)$.  Then, switching to the laser-rotating frame  $\tilde{c}(t) = \hat{c}(t)e^{i\omega_L t}$, $\tilde{\sigma}_n (t) = \hat{\sigma}_n(t) e^{i\omega_L t}$ and treating the 1D continuum as a reservoir, we take a Born-Markov type approximation and obtain the Heisenberg-Langevin equation for $\tilde{c}(t)$,
\begin{equation}\label{Seq:c_dot}
    \dot{\tilde{c}}(t) = i \delta_c \tilde{c}(t) -\frac{\varkappa}{2} \tilde{c}(t)-i g \sum_n \tilde{\sigma}_n +\hat{E}(t)- i \Omega_L.
\end{equation}
Here we introduced the laser detuning from the cavity $\delta_c = \omega_L-\omega_c$ and vacuum noise of the reservoir $\hat{E}(t) =  -i g_c^* \sum_k \hat{b}_k(0) e^{-i(\omega_k-\omega_L) t}$, $\expval{\hat{E}(t)} = 0$, with the time correlator
\begin{equation}
    \expval{\hat{E}(t)\hat{E}^\dagger(t') }  = \varkappa \delta(t-t'),
\end{equation}
where $\varkappa = 2 \abs{g_c}^2 L/c$ and $L$ is  the quantization length of the 1D continuum.
We work in the fast cavity limit, assuming $\varkappa$ is much faster than the timescale of variations in $\tilde{\sigma}_n$. This allows us to adiabatically eliminate the cavity mode, which can be effectively achieved by solving Eq.~\eqref{Seq:c_dot} in the steady state
\begin{equation}\label{Seq:c_t}
    \tilde{c}(t) = \frac{i\Omega_L + i g \sum_n \tilde{\sigma}_n(t) - \hat{E}(t)}{i\delta -\varkappa/2},
\end{equation}
and insert this solution, together with that for the free-space modes together with that for the free-space modes $\hat{a}_{\bm k}$ into the Heisenberg equation of motion of the Pauli lowering operator of an atom $n$
 \begin{align}\label{Seq:dsigma_n_t}
    \dot{\tilde{\sigma}}_n &= i \delta \tilde{\sigma}_n+i \Delta \sigma_n^z(t) \sum_{m} \tilde{\sigma}_m(t) \nonumber\\
    &\quad +\frac{\gamma_c}{2}\hat{\sigma}_n^z(t) \sum_{m}\tilde{\sigma}_m(t)+\hat{\sigma}_n^z(t) \sum_m D(\bm r_n-\bm r_m) \tilde{\sigma}_m(t) \nonumber\\
    &\quad +i \Omega\hat{\sigma}_n^z(t)+  [\hat{f}(t)+\hat{\eta}_n(t)]\hat{\sigma}_n^z(t).
\end{align}
Here we introduced the effective drive Rabi frequency $\Omega$, atom-laser detuning $\delta$, emission rate via the cavity mode $\gamma_c$ and collective shift $\Delta$ that describes the resonant dipole-dipole interaction between pairs of atoms
\begin{align}\label{Seq:Params}
    \Omega &= \frac{i g \Omega_L}{i\delta_c-\varkappa/2}, \quad \delta = \omega_L -\omega_0, \nonumber\\
    \gamma_c &= \frac{2 g^2\varkappa}{2\delta_c^2+\varkappa^2/2}, \quad \Delta = \frac{\delta_c g^2}{\delta_c^2+\varkappa^2/4}.
\end{align}
The vacuum noise $\hat{f}(t)$ in Eq.~\eqref{Seq:dsigma_n_t} is a vacuum noise of the 1D modes reservoir $\hat{E}(t)$ filtred by the cavity:
\begin{gather}\label{Seq:f_def}
    \hat{f}(t) = -\frac{ig}{i\delta_c-\varkappa/2}\hat{E}(t), \\ \notag \expval{\hat{f}(t)\hat{f}^\dagger(t')} = \frac{g^2}{\delta_c^2+\varkappa^2/4}\expval{\hat{E}(t)\hat{E}^\dagger(t')} = \gamma_c\delta(t-t').
\end{gather}
The assumed fast cavity regime is realized when the cavity decay rate is much faster than the atom emission rate through the cavity $\gamma_{c}\ll \varkappa \; \Rightarrow \; 4g^2 \ll \varkappa^2 $ at $\delta_c = 0$.

The elimination of the free-space modes $\hat{a}_{\bm k,\mu}$ was also performed here within a Born-Markov type approximation, yielding the dipole-dipole interaction $D(\bm r_n-\bm r_m)$ mediated by the free-space field modes between different atoms, proportional to the photon Green's function~\cite{solomons_multichannel_2023}. For inter-atomic distances exceeding the optical wavelength, $\abs{\bm r_n-\bm r_m}\gg \lambda = 2\pi c/\omega_L$, this coupling becomes negligible, and we approximate
$D(\bm r_n-\bm r_m) \approx \delta_{nm}\gamma_s/2$, yielding an individual-atom decay at the rate $\gamma_s$ of spontaneous emission to free-space. The dissipation induced by the  reservoir of modes $\hat{a}_{\bm k,\mu}$ is accompanied by the vacuum-field Langevin noise
\begin{align}\label{Seq:xi_n_def}
    \hat{\eta}_n(t) &= i\sum_{\bm k,\mu} g_{\bm k, \mu,n}\hat{a}_{\bm k,\mu}(0) e^{-i(\omega_{\bm k}-\omega_L) t}, \nonumber\\
    \expval{\hat{\eta}_n(t) \hat{\eta}_m^\dagger(t')} &\approx \gamma_s \delta_{nm}\delta(t-t').
\end{align}
Finally, taking resonant conditions  $\delta = \delta_c = 0$, we obtain the Heisenberg-Langevin equation
\begin{align}\label{Seq:sigma_HL}
     \dot{\tilde{\sigma}}_n(t) &=-\frac{\gamma_s}{2}\tilde{\sigma}_n(t)+ i \Omega \hat{\sigma}_n^z(t) \\
     &\quad +\frac{\gamma_c}{2}\hat{\sigma}_n^z(t) \sum_{m}\tilde{\sigma}_m(t)+\hat{\sigma}_n^z(t)[\hat{f}(t)+\hat{\eta}_n(t)]. \nonumber
\end{align}
A similar procedure leads to the Heiseberg equation for $\hat{\sigma}_n^z$, obtaining
\begin{gather}\label{Seq:sigmanz_HL}
	\dot{\hat{\sigma}}_n^z(t) = - \gamma_s(1+\hat{\sigma}_n^z(t))-\\  -2\qty[\tilde{\sigma}_n^\dagger(t)\qty(\hat{f}(t)+\hat{\eta}_n(t)+i\Omega+{\gamma_c\over 2}\sum_{m}  \tilde{\sigma}_m(t))+\text{H.c.}]. \nonumber
\end{gather}
The Heisenberg-Langevin equations \eqref{Seq:sigma_HL} and \eqref{Seq:sigmanz_HL} are equivalent to the density matrix equation [Eq.~\eqref{Eq:ME} in the main text]. It could be rewritten as
\begin{equation}\label{Seq:For_J}
    \dot{\rho}= -i\qty({\cal H}_{\text{eff}}\rho(t)-\rho(t){\cal H}^\dagger_{\text{eff}})+\sum_{i = 1}^{N+1} {\cal C}_i \rho(t) {\cal C}_i^\dagger,
\end{equation}
where we introduce the non-Hermitian Hamiltonian ${\cal H}_{\text{eff}} = 2\Omega \hat{S}_x-\frac{i}{2}\sum_{i = 1}^{N+1}{\cal C}_i^\dagger{\cal C}_i$ and jump operators ${\cal C}  = \{ \sqrt{\gamma_c}\hat{S}_-, \, \sqrt{\gamma_s}\hat{\sigma}_1,...,\sqrt{\gamma_s}\hat{\sigma}_N\}$. Such form is used in Quantum Monte-Carlo trajectories simulations shown in Fig.~\ref{fig:quant_jump} of the main text.

\section{Mean-field results}
\label{app:MF}
The mean-field equation could be obtained either from tracing the master equation [Eq.~\eqref{Eq:ME} of the main text] with the required operator and performing averaging or directly averaging the Heisenberg-Langevin Eqs.~\eqref{Seq:sigma_HL} and~\eqref{Seq:sigmanz_HL}. We define the mean-field values assuming that they are equal for different atoms $\expval{\hat{\sigma}_n} = s_n = s$ and $\expval{\hat{\sigma}_n^z} = s_n^z = s_z$ (with the simplified notation $\tilde{\sigma}_n \rightarrow \hat{\sigma}_n$ from here on). To deal with the multiplications of the operators that exist in Eqs.~\eqref{Seq:sigma_HL}-\eqref{Seq:sigmanz_HL}, we use individual atom factorization
\begin{equation}
	\expval{\hat{\sigma}_n^\dagger \hat{\sigma}_m} = s^* s = \abs{s}^2 \qquad n \neq m,
\end{equation}
that is appropriate when the total momentum $\hat{\bm S}^2 = \hat{S}_x^2+\hat{S}_y^2+\hat{S}_z^2$ is not conserved, as in the case of additional individual decay~\cite{drummond_volterra_1978}.

Following this procedure, we obtain the system of mean-field nonlinear differential equations
\begin{equation}\label{Seq:dif_eq_1}
	\begin{cases}
	\dot{s}_x = -\frac{\gamma}{2}s_x+\frac{\Gamma}{2}s_z s_x,\\
	\dot{s}_y = -\frac{\gamma}{2}s_y-2\Omega s_z +\frac{\Gamma}{2}s_z s_y,\\
	\dot{s}_z =  -\gamma(s_z+1)+2\Omega s_y-\frac{\Gamma}{2}(s_x^2+s_y^2),
	\end{cases}
\end{equation}
where we defined $x,y$ components as $s_x = s+s^*$, $s_y = i(s-s^*)$ and took the effective Rabi drive frequency $\Omega$ to be real without loss of generality. The obtained effective mean-field parameters are related to the model parameters $\gamma_c$, $\gamma_s$  and $N$ by Eq.~\eqref{eq:gammaneq0} in the main text.

From the system of equations~\eqref{Seq:dif_eq_1} one can see that the total spin $\langle\hat{\bm S}^2\rangle_{\rm MF} = N^2(s_x^2+s_y^2+s_z^2)/4$ is indeed not conserved
\begin{equation}
	\dv{\langle \hat{\bm S}^2 \rangle_{\rm MF}}{t}= -{\gamma N^2 \over 2} \qty(s_z^2+s_z+\frac{s_x^2+s_y^2}{2}),
\end{equation}
and, therefore, utilization of the individual atom factorization is consistent.
\begin{figure*}[ht!]
\centering
    \includegraphics[width=\linewidth]{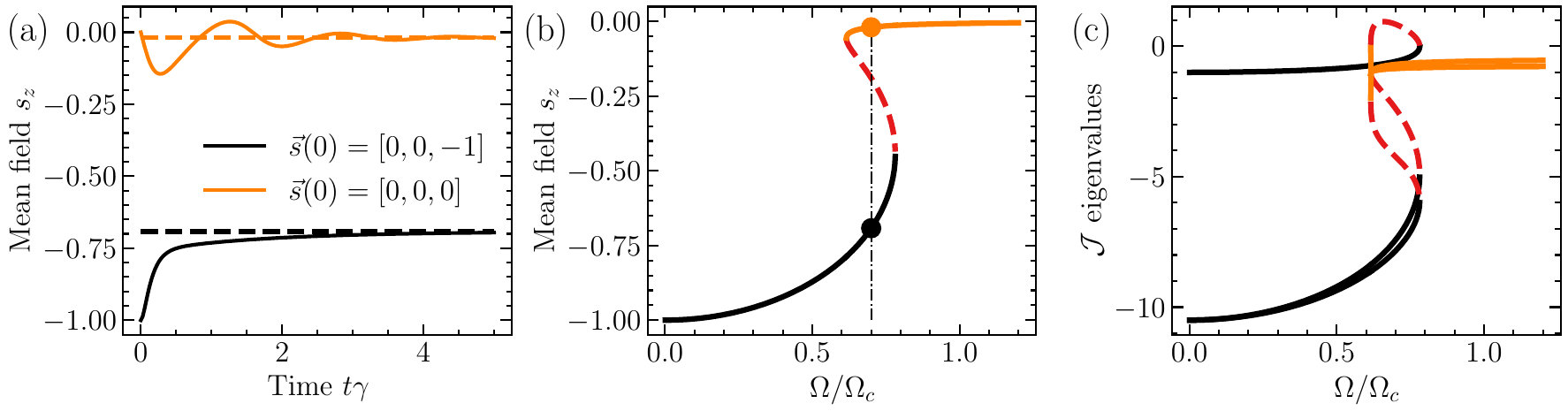}
    \caption{(a) Numerical time solution of the system~\eqref{Seq:dif_eq_1} for different initial conditions at $\Omega/\Omega_c = 0.7$ and $\Gamma/\gamma = 20$. (b) The steady states of the system~\eqref{Seq:dif_eq_1} as function of $\Omega/\Omega_c$ for  $\Gamma/\gamma = 20$. Black and orange solid curves are stable lower and upper branches, and the red dashed curve is an unstable middle branch. (c) Calculated eigenvalues of the Jacobian~\eqref{Seq:Jacob} for each branch in panel (b).}
    \label{fig:MFTime}
\end{figure*}

In the main text, we compare the results of the mean-field and the master equation~\eqref{Eq:ME} with the Dicke $\hat{\bm S}^2$-conserved model with $\gamma_s = 0$
extensively studied before~\cite{drummond_volterra_1978,somech_quantum_2024}. In this case, the mean-field solution that is verified by the exact master equation solution~\cite{somech_quantum_2024} is obtained using collective products factorization~\cite{drummond_volterra_1978} $$\expval{\hat{S}\hat{S_z}} = \expval{\hat{S}}\expval{\hat{S}_z}.$$

Such factorization will lead to the same equations~\eqref{Seq:dif_eq_1} with the relation to the master equation parameters $\gamma = 0$ and $\Gamma \rightarrow \Gamma_c = N \gamma_c$. That factorization is consistent with $\hat{\bm S^2}$-invariance and gives unique physical steady state~\cite{somech_heisenberg-langevin_2023}
\begin{equation}\label{Seq:Dicke_SS}
    s_x = 0, \qquad s_y = \frac{\Omega}{\Omega_c}, \qquad s_z = -\sqrt{1-\frac{\Omega^2}{\Omega_c^2}}
\end{equation}
for $\Omega\leq \Omega_c$ where $\Omega_c = \Gamma_c/4$ is a critical value for this model. In the main text, to compare the two models, we use Eq.~\eqref{Seq:Dicke_SS} with $\Gamma_c = \Gamma$ given by Eq.~\eqref{eq:gammaneq0} of the main text assuming that the number of atoms is large.

Calculating the discriminant of the equation for the steady state Eq.~\eqref{eq:gammaneq0}  we find that depending on the drive Rabi frequency $\Omega$, this polynomial has three real solutions in the region
\begin{align}\label{Seq:bist_reg}
    \frac{1}{2}\sqrt{1+\frac{20\gamma}{\Gamma} -\frac{8\gamma^2}{\Gamma^2}-\sqrt{{\qty(1-\frac{8\gamma}{\Gamma})}^3}} &< \frac{\Omega}{\Omega_c} \nonumber \\
    < \frac{1}{2}\sqrt{1+\frac{20\gamma}{\Gamma} -\frac{8\gamma^2}{\Gamma^2}+\sqrt{{\qty(1-\frac{8\gamma}{\Gamma})}^3}}.
\end{align}

We could obtain approximate analytical solutions for $s_z$ from Eq.~\eqref{eq:gammaneq0} using perturbation theory. We write the perturbative solution of $s_z$ in the small parameter $\gamma/\Gamma$
\begin{equation}\label{Seq:expan}
	s_z = s_z^{(0)}+s_z^{(1)}\gamma +s_z^{(2)}\gamma^2 + {\cal O}(\gamma^3)
\end{equation}
and then substitute it into Eq.~\eqref{eq:gammaneq0} equating the coefficients between $\gamma$. In the zeroth order, we have
\begin{equation}
	8 s_z^{(0)}\Omega^2+16 (s_z^{(0)})^2+16 (s_z^{(0)})^3 \Omega_c^2 = 0 .
\end{equation}
Since the equation is cubic, we will have three solutions that we denote as (a), (b), and (c) in Eq.~\eqref{eq:sz_an} of the main text.

One can see that the solutions (a) and (b) are real only if $\Omega<\sqrt{2}\Omega_c$, thus imposing restrictions on the perturbative solution in all orders. Substituting Eq.~\eqref{Seq:expan} up to the first order $\gamma$ to Eq.~\eqref{eq:gammaneq0} we get
\begin{align}
    -2 s_z^{(0)}\Gamma-2(s_z^{(0)})^2 \Gamma + 8 s_z^{(1)} \Omega^2+ \nonumber\\
    32 s_z^{(0)} s_z^{(1)} \Omega_c^2+48 (s_z^{(0)})^2 s_z^{(1)} \Omega_c^2 = 0.
\end{align}
After the substitution of $s_z^{(0)}$ from the previous step we find $s_z^{(1)}$
\begin{equation}
	s_z^{(1,c)} = 0,\quad s_z^{(1,^a_b)} = \frac{\Gamma}{16 \Omega_c^2}\qty(1\mp \sqrt{\frac{1}{1-2\Omega^2/\Omega_c^2}}).
\end{equation}
Repeating this procedure one more time, we get second-order corrections:
\begin{subequations}
\begin{gather}
	s_z^{(2,^a_b)} = \pm \frac{\left(\sqrt{\Omega_c^2-2 \Omega ^2}\mp\Omega_c \right)}{ \left(\Omega_c
   \left(\pm\sqrt{\Omega_c^2-2
   \Omega ^2}+\Omega_c\right)-2 \Omega ^2\right)}\times \\\times \frac{\left(\Gamma ^2
   \left(\Omega_c
   \left(\pm\sqrt{\Omega_c^2-2
   \Omega ^2}+\Omega_c\right)-\Omega ^2\right)+16
   \Omega ^2 \Omega_c^2-8
   \Omega_c^4\right)}{128
   \Omega_c^3
   \left(\Omega_c^2-2 \Omega
   ^2\right)},\notag
   \end{gather}
\begin{equation}
   s_z^{(2,c)} = -\frac{1}{8\Omega^2}.
\end{equation}
\end{subequations}

One can see that $s_z^{(2,c)}$ diverges for small $\Omega$, which is in contradiction with expansion~\eqref{Seq:expan}, so this solution works only for big enough values of $\Omega$ compared with $\gamma$.
One can see that $s_z^{(2,c)}$ diverges for small $\Omega$, which is in contradiction with expansion~\eqref{Seq:expan}, so this solution works only for big enough values of $\Omega$ compared with $\gamma$.

The stability analysis of these solutions could be performed by calculating the Jacobian matrix ${\cal J}$. From the system of differential equation~\eqref{Seq:dif_eq_1} we get
\begin{equation}\label{Seq:Jacob}
	{\cal J} = \mqty[-\frac{\gamma}{2}+\frac{\Gamma}{2}s_z & 0 & \frac{\Gamma}{2}s_x \\ 0 & -\frac{\gamma}{2}+\frac{\Gamma}{2}s_z & -2\Omega +\frac{\Gamma}{2}s_y \\ -\Gamma s_x & 2\Omega-\Gamma s_y & -\gamma].
\end{equation}

The steady-state point for each $\Omega$ is stable if and only if all eigenvalues of the Jacobian matrix ${\cal J}$ with substituted steady-state point have real parts that are negative~\cite{hirsch_differential_1974}. For the system~\eqref{Seq:dif_eq_1}, it turns out that the lower and upper branches of the mean-field steady-state solution are stable [see Fig.~\eqref{fig:MFTime}], whereas the middle branch always has one eigenvalue with a positive real part and therefore is unstable.

In the limit case $\Gamma\gg \gamma$, we can show the instability of the middle branch in Fig.~\ref{fig:MFTime}(b) analytically: for the solution (c) from Eq.~\eqref{eq:sz_an} the eigenvalues of~\eqref{Seq:Jacob}
$$l_{1} = -\gamma/2 \qquad l_{2,3} = -\frac{1}{4}\qty(3\gamma \pm \sqrt{\gamma^2-64\Omega^2}),$$
always have a negative real part. For the solutions (a) and (b) form Eq.~\eqref{eq:sz_an} we substitute in~\eqref{Seq:Jacob} and get
\begin{align}
l_{1} &= -\frac{\gamma}{2}+s_z \Gamma,\qquad l_{2} = \frac{s_z \Gamma}{2} -\gamma\qty(\frac{1}{2}+\frac{8\Omega^2}{s_z^2 \Gamma^2})+{\cal O}(\gamma^2), \nonumber \\
l_3 &= -\gamma\qty(1-\frac{8\Omega^2}{s_z^2\Gamma^2})+{\cal O}(\gamma^2).
\end{align}
Eigenvalues $l_{1,2}$ are always negative whereas $l_3$ is positive if we substitute $s_z^{(b)}$ from Eq.~\eqref{eq:sz_an} of the main text and negative for $s_z^{(a)}$. So the middle branch is unstable for any value of $\Gamma/\gamma$ and in the region of $\Omega$ given by Eq.~\eqref{Seq:bist_reg}, the mean-field has only two stable steady-states. In Fig.~\ref{fig:MFTime}(a), we demonstrate that both steady states could be reached depending on the initial conditions.

\section{Estimation of the critical point}
\label{app:PTpoint}

As discussed in the main text, the steady-state solution around the phase transition point $\Omega_{\rm PT}$ is spanned by the $\rho_{\pm}$ decomposition [Eq.~\eqref{eq:rho_ss_ppm}]. Exactly at the phase transition point, we expect the coefficients $a_\pm$ to be equal, leading to
\begin{equation}\label{Seq:PT_point}
    \rho_s \simeq \frac{\rho_+ + \rho_-}{2} \quad \text{at} \quad \Omega = \Omega_{\rm PT}.
\end{equation}
However, the exact numerical diagonalization of the Liouvillian, which is necessary to determine $\Omega_{\rm PT}$, is feasible only up to $N = 18$. As seen in Fig.~\ref{fig:results}(b), in this case, the peaks have sufficient overlap and do not fully cover $\rho_s$ due to the finite small size of the system. Instead, similarly to the Liouvillian gap calculation, we use steady-state values of the observables. Taking the trace with $\hat{S}_z$ in Eq.~\eqref{Seq:PT_point} and utilizing the fact that the two mean-field branches are in strong agreement with $\Tr[\hat{S}_z \rho_\pm]$, we obtain
\begin{equation}\label{Seq:PT_in_Sz}
    \expval{\hat{S}_z} = \frac{S_z^+ + S_z^-}{2} \quad \text{at} \quad \Omega = \Omega_{\rm PT},
\end{equation}
where $S_z^\pm$ are the mean-field values for the lower and upper branches in the bistability region, and $\langle \hat{S}_z \rangle$ is the steady-state solution of the master equation.

\begin{figure}[t!]
    \centering
    \includegraphics[width=1\columnwidth]{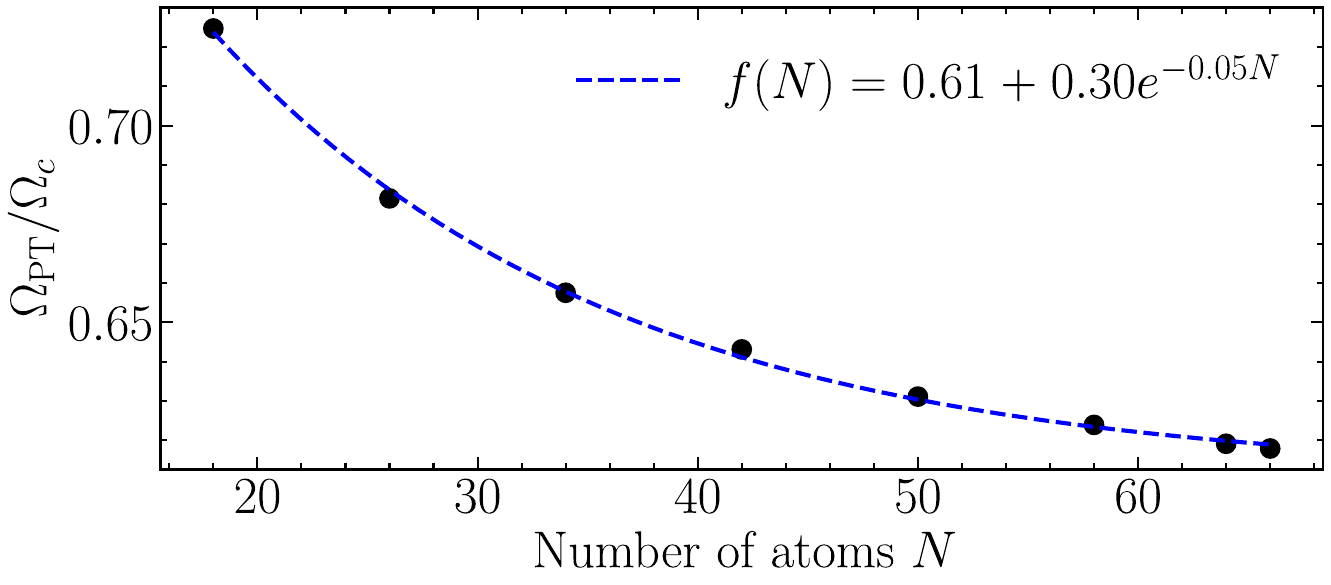}
    \caption{Estimation of the phase transition point from Eq.~\eqref{Seq:PT_point} (black dots). Exponential fit as a function of number of atoms (blue dashed line).}
    \label{app:fig_pt_point}
\end{figure}

We solved Eq.~\eqref{Seq:PT_in_Sz} as a function of $\Omega$ for different values of $N$ and observed an exponential convergence to a constant value, $\Omega \approx 0.61 \Omega_c$, as shown in Fig.~\ref{app:fig_pt_point}. As seen in Fig.~\ref{fig:res1}(a-b), the solution of Eq.~\eqref{Seq:PT_in_Sz} corresponds to the transition point between the two mean-field branches, as well as to the $m$-distribution exhibiting two peaks of equal height. We then demonstrate the closing of the Liouvillian gap in Fig.~\ref{fig:lgap} at the estimated point, proving the occurrence of a first-order dissipative phase transition in our system~\cite{vicentini_critical_2018,minganti_spectral_2018}.

\section{Properties of the totally mixed state}
\label{app:mixed}
In the system with individual decay where all the $2^N$ states in the Hilbert space could be reached, we expect that under string drive the system density matrix reaches a totally mixed state (i.e. proportional to identity matrix). In the basis of the angular momentum states $\ket{j,m}$, this density matrix has a block-diagonal form with $N/2$ blocks corresponding to each $j$ with the degeneracies of each $j$, state~\cite{damanet_cooperative_2016,shammah_superradiance_2017}
\begin{equation}
    d_{j}^N = \frac{N! (2j+1)}{(N/2-j)!(N/2+j+1)!},
\end{equation}
on the diagonals
\begin{gather}\label{Seq:rho_mix}
    \rho_{\rm mix} =  \frac{1}{2^N} \times \\ \mqty(
    \mqty{
        1 & 0& \dots & 0 \\
        0 & 1 & \dots & 0 \\
         &  &   \ddots &  &  \\
        \dots & 0 & 0 & 1
    } & & & & \\
    & \mqty{
        d_{N/2-1}^N & 0 & \dots \\
         & \ddots &  \\
        \dots & 0 & d_{N/2-1}^N
    } & & & \\
    & & \ddots & & \\
    & & & & d_{0}^{N}
). \notag
\end{gather}
Calculating the systems averages with $\rho_{\rm mix}$ gives
\begin{subequations}
\begin{equation}
    \expval{\hat{S}_z} = 0, \qquad  \expval{\hat{S}_z^2} = N/4
\end{equation}
\begin{equation}
    \expval{\hat{S}_+ \hat{S}_-} = N/2
\end{equation}
\begin{equation}
    \expval{\hat{\bm S^2}} = 3N/4
\end{equation}
\begin{equation}
     \expval{\hat{S}_+ \hat{S}_+\hat{S}_- \hat{S}_-} = \frac{(N-1)N}{2}
\end{equation}
\begin{equation}
    g^{(2)} = \frac{ \expval{\hat{S}_+ \hat{S}_+\hat{S}_- \hat{S}_-}}{\expval{\hat{S}_+ \hat{S}_-}^2} = 2\qty(1-\frac{1}{N})
\end{equation}
\end{subequations}
In Fig.~\ref{fig:res1} we show that this asymptotics is reached at $\Omega>\Omega_c$

\section{Decomposition of $\rho_1$ matrix}
\label{app:rho1}

\begin{figure}[ht!]
    \includegraphics[width=0.92\columnwidth]{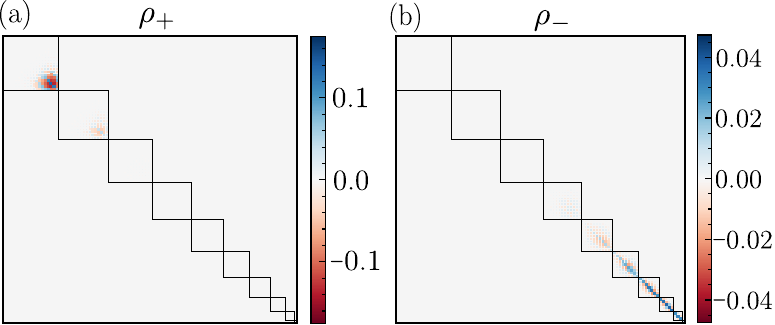}
    \caption{The decomposition of the matrix $\rho_1$ into the $\rho_+$ (right panel) and $\rho_-$ (left panel) for $N = 18$ and $\Omega= 0.73 \Omega_c$. The matrices are shown in the Dicke block-diagonal basis, where the blocks (black squares) correspond to the total spin $j$, decreasing from $j = N/2$ at the top left. See~\cite{shammah_open_2018} for details on the block-diagonal representation.}
    \label{fig:rho_decomp_app}
\end{figure}

The decomposition of the matrix $\rho_1$ into the $\rho_\pm$ components is performed according to the procedure described in Sec.~\ref{sec:bistability} of the main text. The resulting matrices $\rho_\pm$ for $N = 18$ and $\Omega = 0.73 \Omega_c$ are shown in Fig.~\ref{fig:rho_decomp_app} in a block diagonal generalized Dicke states representation described in~\cite{shammah_open_2018}. Their structure verifies the properties discussed in the main text: $\rho_+$ mostly consist of the states with $j = N/2$ (Dicke manifold) where the CRSS state is defined~\cite{somech_quantum_2024}. That is also in consistent with the high fidelity of $\rho_+$ with the CRSS state shown in Fig.~\ref{fig:crss_squeezing}(a) of the main text. On the other hand, density matrix $\rho_-$ is distributed over the states with low $j$ values. In the limit $N \to \infty$ $\rho_-$ is expected to be distributed uniformly (taking into account degeneracies of the states with $j<N/2$) over all $j$ states excluding $j = N/2$ thus closely resembling the totally mixed state [see Eq.~\eqref{Seq:rho_mix}].

We performed the decomposition of $\rho_1$ for different values of $\Omega$ in the bistability region and calculated the coefficients $a_\pm$ using the trace and orthogonality conditions described in Sec.~\ref{sec:bistability} of the main text. For the large system sizes ($N > 18$) exact dioganalization of the Liouvillian is not feasible, so we used shift-inverted method looking only for the Liouvillan eigenvalue corresponding to the $\lambda_1$ which we know from the Fig.~\ref{fig:lgap} of the main text where $\lambda_1$ was obtained from the fit of the long time dynamics of $\langle \hat{S}_z \rangle$ to the steady state. We then calculated the corresponding eigenvector $\rho_1$ and found $a_\pm$ up to $N = 32$.

\begin{figure}[t!]
    \centering
    \includegraphics[width=.99\columnwidth]{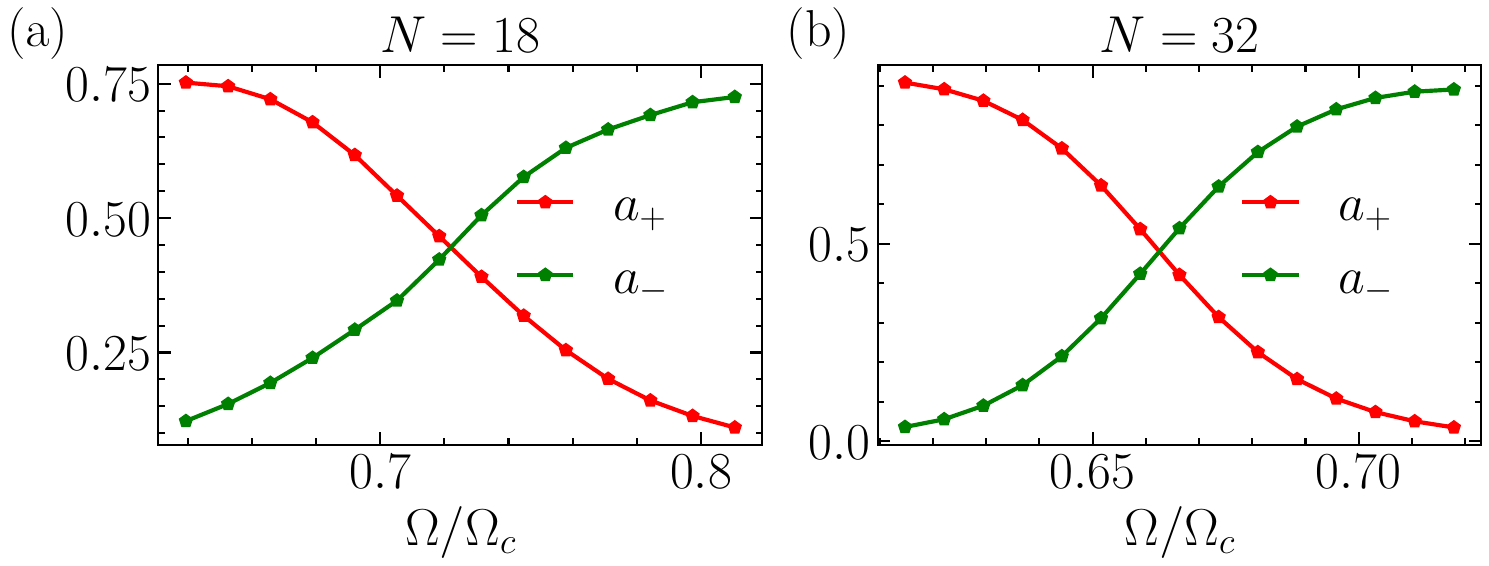}
    \includegraphics[width=.99\columnwidth]{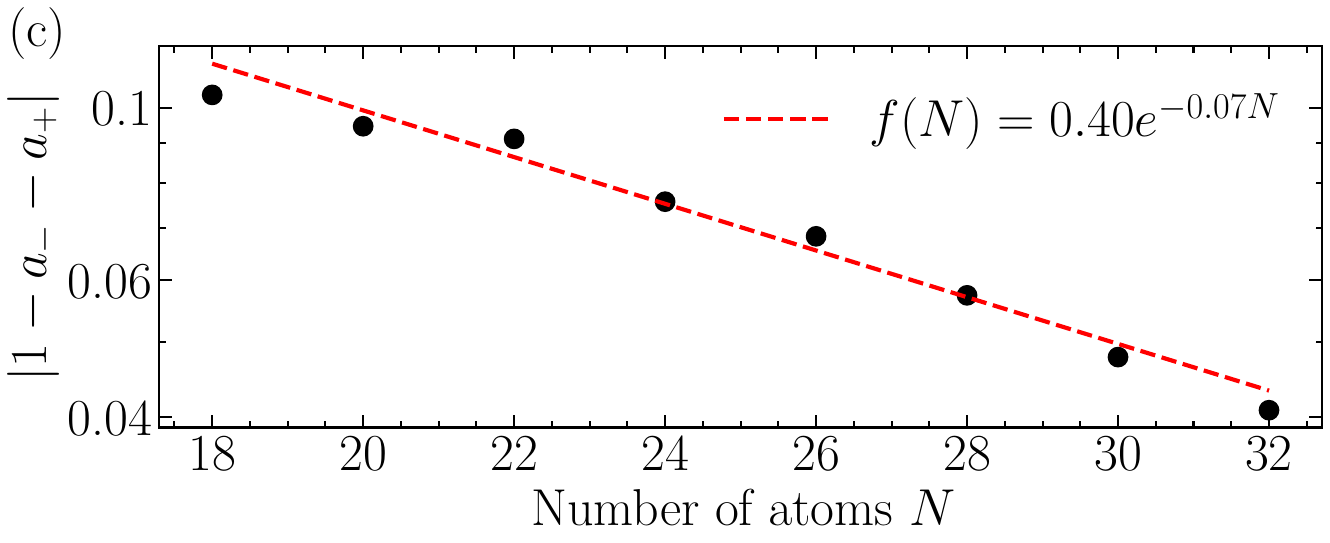}
    \caption{The coefficients $a_\pm$ calculated from the trace and orthogonality conditions as explained in Sec.~\ref{sec:bistability} of the main text as function of $\Omega$ with fixed $\gamma_{c} = 10 \gamma_s$ for (a) $N = 18$ and (b) $N = 32$ in their own bistability regions. (c) The error of the decomposition $|1-a_- - a_+|$ as function of $N$ calculated at the phase transition point $\Omega_{\rm PT}$ for each $N$. Fit (red dashed line) supports an exponential decrease of the error with $N$.}
    \label{fig:apam_err}
\end{figure}

The results of the procedure are shown in Fig.~\ref{fig:apam_err}. One can see that for each $N$ the coefficients $a_{\pm}$ are equal at the point that coincides with the phase transition point $\Omega_{\rm PT}$ estimated in Appendix~\ref{app:PTpoint}. We then calculate the error of the decomposition $|1-a_- - a_+|$ as function of $N$ at $\Omega = \Omega_{\rm PT}$ for each $N$ in Fig.~\ref{fig:apam_err}(c) and observe an exponential decrease of the error with $N$ thus indicating that the decomposition becomes exact in the thermodynamic limit.

\section{Spin squeezing calculation}
\label{app:spin_squeezing}

Following the mean-field analysis, we examine the deviations from the mean-field solution to assess atomic correlations in the steady state. We adopt the analytical approach from \cite{somech_heisenberg-langevin_2023} used for the driven Dicke model with only collective decay and generalize it to the situation where individual decay is also present. Using the Holstein-Primakoff transformation~\cite{ma_quantum_2011, lee_dissipative_2014, somech_heisenberg-langevin_2023}, we establish a connection between spin squeezing and bosonic squeezing and calculate the latter analytically.

\subsection{Holstein-Primakoff transformation}
Consider the mean-field steady state vector $\langle\hat{\bm S}\rangle_{\rm MF} = \frac{N}{2} \mqty[0 & s_y & s_z]^T$. We rotate the coordinate frame around the $x$ axis so that the new axis $z'$ is along the mean-field direction $\langle \bm S'\rangle_{\rm MF} = j_{\rm MF} \mqty[0 & 0 & 1]^T$ where we defined the total mean-field spin $j_{\rm MF} = N\sqrt{s_z^2+s_y^2}/2$. The new primed mean-field components are related to the initial ones as
\begin{equation}\label{Seq:th_def}
	\mqty[s'_x \\ s'_y \\ s'_z] = \mqty[1 & 0 & 0 \\ 0 & \cos\theta & -\sin \theta \\ 0 & \sin \theta & \cos \theta]\mqty[s_x \\ s_y \\ s_z],
\end{equation}
where $\theta = \arccos(s_z/\sqrt{s_z^2+s_y^2})$ and the relations between individual atom operators are given by

\begin{multline}\label{Seq:trans_mat}
	\mqty[\sigma'_n \\ \qty(\sigma'_n)^\dagger \\ \qty(\sigma'_n)^z] = \mqty[\cos^2{\theta\over 2} & \sin^2{\theta \over 2} & \frac{i}{2}\sin \theta\\\sin^2{\theta \over 2} &\cos^2{\theta\over 2} & - \frac{i}{2}\sin \theta \\ i \sin \theta & -i \sin \theta & \cos \theta ]\mqty[\hat{\sigma}_n \\ \hat{\sigma}_n^\dagger \\ \hat{\sigma}_n^z] \quad \Rightarrow \\  \mqty[\hat{\sigma}_n \\ \hat{\sigma}_n^\dagger \\ \hat{\sigma}_n^z]  = \mqty[\cos^2{\theta\over 2} & \sin^2{\theta \over 2} & -\frac{i}{2}\sin \theta\\\sin^2{\theta \over 2} &\cos^2{\theta\over 2} &  \frac{i}{2}\sin \theta \\ - i \sin \theta & i \sin \theta & \cos \theta ]\mqty[\sigma'_n \\ \qty(\sigma'_n)^\dagger \\ \qty(\sigma'_n)^z].
\end{multline}

Within the primed frame, we now use the Holstein-Primakoff transformation to account for small spin fluctuations around the mean spin directed at $z'$. Typically, as in the Dicke problem~\cite{somech_heisenberg-langevin_2023}, the Holstein-Primakoff transformation is perfomed on collective variables:
\begin{subequations}\label{Seq:HP_usual}
\begin{equation}
    {S'}_{+} = \sum_{n}(\sigma'_n)^\dagger \rightarrow \sqrt{2j_{\rm MF}}\hat{b} +{\cal O}\qty(\frac{1}{j_{\rm MF}}),
\end{equation}
\begin{equation}
    S'_- = \sum_{n}\sigma'_n \rightarrow \sqrt{2j_{\rm MF}}\hat{b}^\dagger +{\cal O}\qty(\frac{1}{j_{\rm MF}}),
\end{equation}
\begin{equation}\label{Seq:Sz_boson}
    S'_z \rightarrow j_{\rm MF}+{\cal O}\qty(\frac{1}{j_{\rm MF}}).
\end{equation}
\end{subequations}
Here $\hat{b}$ and $\hat{b}^\dagger$ are bosonic ladder operators, satisfying $[\hat{b},\hat{b}^\dagger] = 1$, that represent small spin fluctuations for $j_{\rm MF}\gg 1$. In our case, where individual decay $\gamma_s$ and related noise $\hat{\eta}_n$ for each atom are present, the summation of Eqs.~\eqref{Seq:sigma_HL}-\eqref{Seq:sigmanz_HL} could not be rewritten in terms of collective operators used in Eqs.~\eqref{Seq:HP_usual}. Therefore, to describe our system using bosonic operators, we must introduce the local bosonic operators for each atom $n$ $\hat{a}_n$ and $\hat{a}_n^\dagger$ with $[\hat{a}_n,\hat{a}_m^\dagger] = \delta_{nm}$. Based on Eq.~\eqref{Seq:Sz_boson} we substitute
\begin{subequations}
\label{Seq:HP_trans}
    \begin{equation}
        (\sigma'_n)^z \rightarrow 2j_{\rm MF}/N,
    \end{equation}
and then, for the commutation relation $[(\sigma'_n)^\dagger, \sigma'_n]=(\sigma'_n)^z$ to hold, we set
    \begin{equation}
        \sigma'_n = \sqrt{2j_{\rm MF}/N}\hat{a}_n^\dagger,\qquad (\sigma'_n)^\dagger  = \sqrt{2j_{\rm MF}/N} \hat{a}_n.
    \end{equation}
\end{subequations}
To find an equation of motion for the spin fluctuations $\hat{a}_n$ we use~\eqref{Seq:trans_mat}
\begin{equation}
	\dv{(\sigma'_n)^\dagger}{t} = \sin^2{\theta\over 2}\dot{\hat{\sigma}}_n+\cos^2{\theta\over 2}\dot{\hat{\sigma}}_n^\dagger-\frac{i}{2}\sin\theta\dot{\hat{\sigma}}_n^z,
\end{equation}
and then substitute Eqs.~\eqref{Seq:sigma_HL}-\eqref{Seq:sigmanz_HL}. Then, we apply the transformation~\eqref{Seq:HP_trans} and neglect terms of second order in the fluctuations, $\hat{a}_n \hat{\xi}_{n}$, $\hat{a}_n^\dagger \hat{\xi}_n$, $\hat{a}_n \hat{a}_m$, $\hat{a}_n \hat{a}_m^\dagger$, $\hat{a}_n \hat{f}$, $\hat{a}_n^\dagger \hat{f}$, obtaining the linearized Heisenberg-Langevin equation
\begin{align}\label{Seq:dan}
    \dot{\hat{a}}_n &=\frac{\gamma_cj_{\rm MF}}{N} \cos \theta \sum_{m\neq n} \hat{a}_m -\frac{\gamma}{2}\hat{a}_n-\frac{\gamma}{4}\sin^2\theta (\hat{a}_n -\hat{a}_n^\dagger) \nonumber \\
    &\quad +\sqrt{\frac{j_{\rm MF}}{2N}}\left[\cos\theta(\hat{f}+\hat{f}^\dagger+\hat{\eta}_n+\hat{\eta}_n^\dagger)\right. \nonumber \\
    &\quad \left.+(\hat{f}^\dagger-\hat{f}+\hat{\eta}_n^\dagger-\hat{\eta}_n)\right].
\end{align}
Next, we introduce the collective bosonic operator $\hat{a} = \frac{1}{\sqrt{N}}\sum_n \hat{a}_n$ and sum Eq.~\eqref{Seq:dan} over $n$, finding
\begin{gather}\label{Seq:a_eqs}
    \dot{\hat{a}} =\frac{\Gamma j_{\rm MF} \cos \theta}{N}\hat{a} -{\gamma\over 2} \hat{a}-{\gamma\over 4}\sin^2\theta (\hat{a} -\hat{a}^\dagger)\\+\sqrt{j_{\rm MF}\over 2N}\qty[\cos\theta(\hat{F}+\hat{F}^\dagger+\hat{\eta}+\hat{\eta}^\dagger)+(\hat{F}^\dagger-\hat{F}+\hat{\eta}^\dagger-\hat{\eta})],\notag
\end{gather}
where $\hat{\eta} =\frac{1}{\sqrt{N}} \sum_n \hat{\eta}_n$ and $\hat{F} = \sqrt{N} \hat{f}$. Finally, solving Eq.~\eqref{Seq:a_eqs} as a system of equations for $\hat{a}$ and $\hat{a}^\dagger$ in steady state, $t \gg [\Gamma \cos(\theta)+\gamma]^{-1}$, we obtain
\begin{widetext}
    \begin{align}\label{Seq:a_t}
    \hat{a}(t) &= \sqrt{j_{\rm MF}\over 2N}\int\limits_0^t \dd{t'}\exp[\qty(-\Gamma j_{\rm  MF} \cos\theta/N+\frac{\gamma}{2})(t'-t)] \left\{ \qty(\hat{F}(t')+\hat{\eta}(t'))\qty(\cos\theta-\exp[{\gamma\over 2}\sin^2\theta(t'-t)])\right. \nonumber \\
&\quad \left.+\qty(\hat{F}^\dagger(t')+\hat{\eta}^\dagger(t'))\qty(\cos\theta+\exp[{\gamma\over 2}\sin^2\theta(t'-t)])\right\}.
    \end{align}
\end{widetext}

\subsection{Spin squeezing}
The spin squeezing parameter is given by  $ \xi^2 = \min_\varphi \text{Var}[\hat{S}'_\varphi]N/|\langle \hat{\bm S} \rangle|^2$, recalling that $\hat{S}'_\varphi$ is the spin vector component perpendicular to the mean-spin direction. Therefore, in terms of the primed frame from Eqs.~\eqref{Seq:trans_mat} and \eqref{Seq:HP_trans}, we have
\begin{equation}
    \hat{S}'_\varphi
    = \sqrt{j_{\rm MF}\over 2N}(e^{i\varphi}\hat{a}^\dagger+e^{-i\varphi}\hat{a}).
\end{equation}
Moreover, considering the mean spin $|\langle \hat{\bm S} \rangle|^2 = N^2(s_z^2+s_y^2)/4 = j_{\rm  MF}^2$, the spin squeezing parameter is then simply given by
\begin{equation}\label{Seq:xi_2_def}
    \xi^2 = \min_\varphi \text{Var}[e^{i\varphi}\hat{a}^\dagger+e^{-i\varphi}\hat{a}] = 1+2\expval{\hat{a}^\dagger\hat{a}}-2\abs{\expval{\hat{a}}^2}.
\end{equation}
The required correlators of $\hat{a}$ are then obtained from the solution~\eqref{Seq:a_t} as
\begin{widetext}
\begin{subequations}\label{Seq:sq_ans}
\begin{equation}
	\expval{\hat{a}^\dagger \hat{a}} = \frac{(\gamma+\Gamma)\abs{\bm s}}{4}\qty[\frac{1}{-\Gamma\abs{\bm s}\cos\theta+\gamma(1+\sin^2\theta)}+\frac{2\cos\theta}{-\Gamma \abs{\bm s} \cos\theta+\gamma(1+\sin^2\theta/2)}+\frac{\cos^2\theta}{-\Gamma \abs{\bm s}\cos \theta+\gamma}],
\end{equation}
\begin{equation}
	\expval{\hat{a}^2} = \frac{(\gamma+\Gamma)\abs{\bm s}}{4}\qty[\frac{\cos^2\theta}{-\Gamma \abs{\bm s}\cos\theta+\gamma}-\frac{1}{-\Gamma \abs{\bm s} \cos\theta+\gamma(\sin^2\theta+1)}],
\end{equation}
\end{subequations}
\end{widetext}
where we introduced $\abs{\bm s} = 2j_{\rm MF}/N = \sqrt{s_z^2+s_y^2}$ and used the Langevin-noise correlators derived from Eqs.~\eqref{Seq:f_def} and~\eqref{Seq:xi_n_def}
\begin{subequations}
    \begin{equation}
	\expval{\hat{F}(t')\hat{F}^\dagger(t'')} = N\gamma_c \delta(t'-t'')
\end{equation}
\begin{equation}
    \expval{\eta(t')\eta^\dagger(t'')} = \gamma_s\delta(t'-t'').
\end{equation}
\end{subequations}

Now, we consider spin squeezing in some limiting cases. First of all, if in Eqs.~\eqref{Seq:sq_ans} we set $\gamma$ to be equal zero and substitute the solution Eq.~\eqref{Seq:Dicke_SS}, we get the squeezing in the Dicke model for $N \to \infty$ ~\cite{lee_dissipative_2014,somech_quantum_2024,somech_heisenberg-langevin_2023}
\begin{equation}
    \xi^2 = -\cos\theta = \sqrt{1-\frac{\Omega^2}{\Omega_c^2}},
\end{equation}
where we found $\theta$ from the analytical expression~\eqref{Seq:Dicke_SS}.

\begin{figure}[t!]
    \centering
    \includegraphics[width=\linewidth]{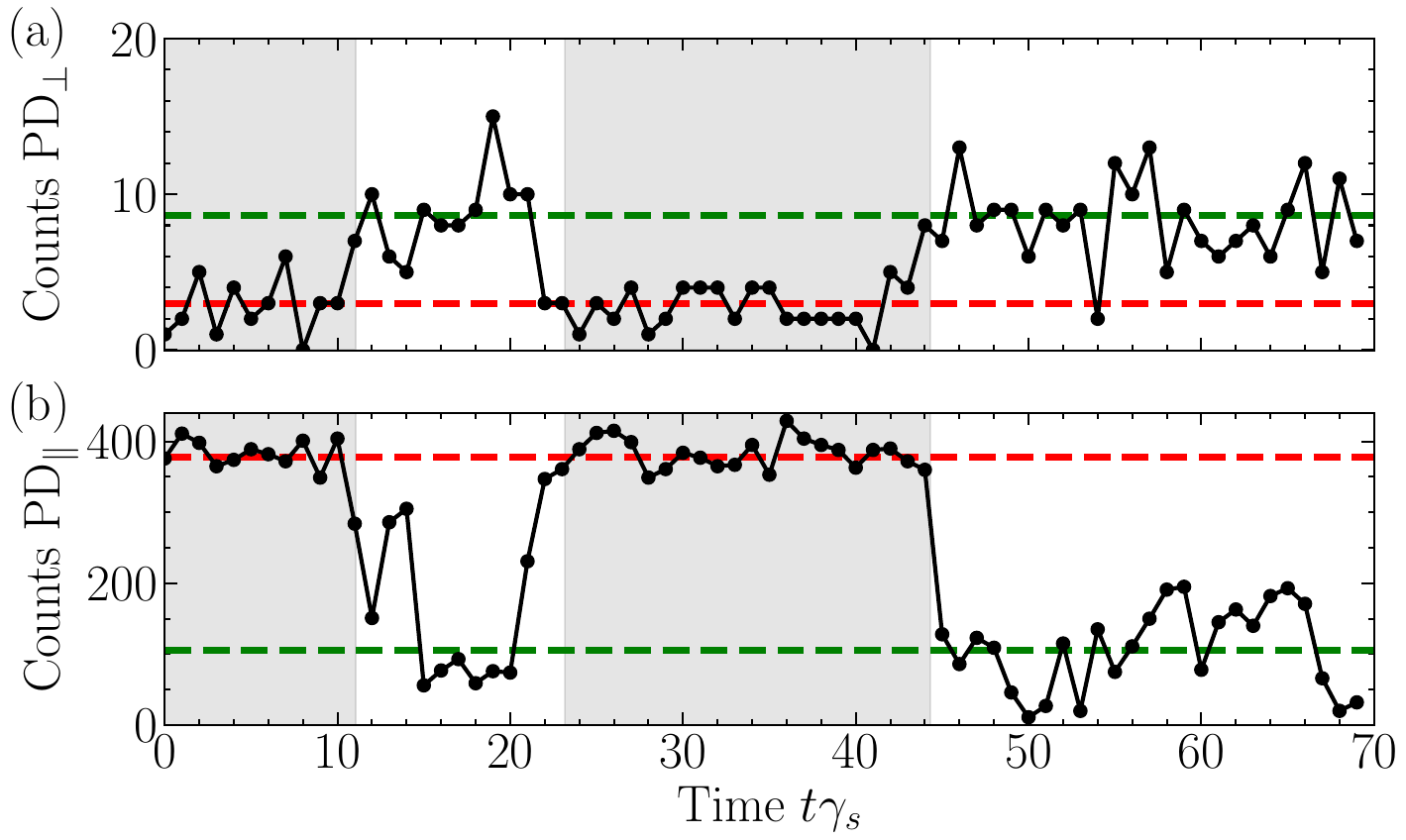}
    \caption{Photon count trajectories with a bin size $\gamma_s \Delta t = 1$ for $N = 18$, $\gamma_{c} = 10\gamma_s$ and $\Omega = 0.73\Omega_c$ (a) Counts of the individual jumps with operators $\{\, \sqrt{\gamma_s}\hat{\sigma}_1,...,\sqrt{\gamma_s}\hat{\sigma}_N\}$  (black dots) exhibit switching between two values obtained from the mean-field steady state $s_z$ for the same drive substitution to the Eq.~\eqref{eq:PDperp} (red and green dashed lines). (b) Counts of the collective jumps with the operator $\sqrt{\gamma_{c}}\hat{S}_-$  (black dots) also exhibit switching between two values obtained from the Eq.~\eqref{eq:PDpar} after substitution $\text{Tr}[\hat{S}_+\hat{S}_-\rho_\pm]$ for corresponding drive value (red and green dashed lines). The jump counting are performed at the same trajectory that is used in Fig.~\ref{fig:quant_jump} }
    \label{fig:photon_count}
\end{figure}

When $\gamma_s$ is considered, we have two stable branches in the bistable region shown in Fig.~\ref{fig:MFTime}(b). However, Eq.~\eqref{Seq:sq_ans} is derived assuming that $j_{\rm MF}\gg 1$, which holds only for the lower branch [black solid line in Fig.~\ref{fig:MFTime}(b)]. For example, the spin squeezing obtained by substiuting into Eqs.~\eqref{Seq:xi_2_def} and~\eqref{Seq:sq_ans} the lower-branch mean-field solution obtained numerically from Eq.~\eqref{eq:gammaneq0} for different number of atoms is shown in Fig.~\ref{fig:res1}(d) of the main text and exhibits excellent agreement with the exact numerical result, obtained directly from the density matrix solution of the master equation~\eqref{Eq:ME}. This holds for the relevant regime wherein $\langle \hat{S_z} \rangle$ coincides with the mean-field lower branch solution. More insights follow from the comparison with the squeezing calculated on $\rho_{+}$, corresponding to the mean-field lower branch density matrix from the spectral decomposition. It agrees with analytics almost until the end of the bistability region [Fig.~\ref{fig:crss_squeezing}(b) in the main text].

In the limiting case $\Gamma \gg \gamma \neq 0$ we still have $\xi^2 = -\cos\theta$, but now we substitute $\theta$ from $s_z^{(0,a)}$ from Eq.~\eqref{eq:sz_an}:
\begin{equation}
    \xi^2 = \frac{1+\sqrt{1-2\Omega^2/\Omega_c^2}}{\sqrt{2}\sqrt{1+\Omega^2/\Omega_c^2+\sqrt{1-2\Omega^2/\Omega_c^2}}}.
\end{equation}
This result is valid in the region $\Omega \leq \Omega_c/\sqrt{2}$  wherein the lower-branch solution is relevant. Since the squeezing parameter $\xi^2$ decreases with $\Omega$ it thus obtains its optimal (minimal) value $1/\sqrt{3}$ at $\Omega = \Omega_c/\sqrt{2}$.

\section{Quantum jumps statistics}
\label{app:photon_counting}

In experiments, the quantum trajectories shown in the main text emerge by the record of photon-counting measurements~\cite{ferioli_non-equilibrium_2023}, as schematically illustrated in Fig.~\ref{fig:sketch}(a). We now show how this works for the observables the $\hat{J}_z$ and $\hat{J}_+\hat{J}_-$ which are related to the direct photon-detection of the individual and collective dissipation ports ($\hat{J}_{x,y}$ can be extracted from homodyne detection on the collective dissipation port). To this end, we keep the time record of emitted photons in both ports for the same quantum trajectory simulations used in the main text. The algorithm provided by QuTiP~\cite{johansson_qutip_2013} records the time of each jump operator's action, allowing one to compute the number of jumps within each selected time bin~\cite{cabot_quantum_2023}.

\begin{figure*}[t!]
    \centering
    \includegraphics[width=1\linewidth]{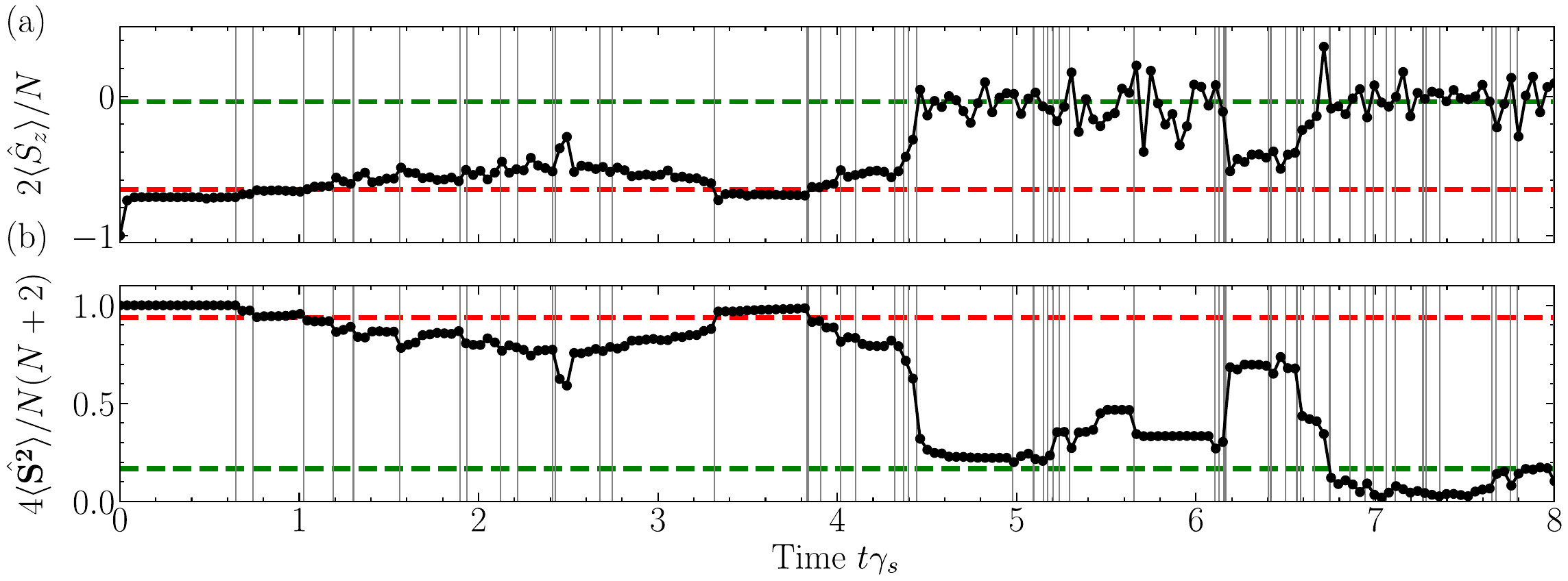}
    \caption{Quantum switching between stable states $\rho_{\pm}$ manifests in quantum-trajectory simulations, performed using QuTip~\cite{johansson_qutip_2013} for $N = 18$, $\gamma_c = 10\gamma_s$, and $\Omega = 0.73\Omega_c$, with all atoms initially in the ground state. Gray vertical lines show the action of the individual jump with one of the operators from $\{\, \sqrt{\gamma_s}\hat{\sigma}_1,...,\sqrt{\gamma_s}\hat{\sigma}_N\}$
(a) Magnetization $\langle \hat{S}_z \rangle$ in a single trajectory exhibits switching between the two mean-field values, corresponding to $\rho_{\pm}$ (red and green dashed curves).
(b) Total angular momentum for the same trajectory compared to $\text{Tr}[\hat{\bm S}^2 \rho_+]$ and $\text{Tr}[\hat{\bm S}^2 \rho_-]$ at the same $\Omega$ (red and green dashed curves, respectively).}
    \label{fig:short_jump}
\end{figure*}

The detection probability of individual quantum jumps by the photodetector labeled PD$_{\perp}$ is proportional to the population of each atom, given by $(1 + \hat{\sigma}_n^z)/2$. Therefore, the total number of individual jumps provides information about the observable $\hat{S}_z$, as expressed by the equation
\begin{equation}\label{eq:PDperp}
    \text{Counts PD}_{\perp}(t\,|\, t+\Delta t) = \left(\frac{N}{2} + \langle \hat{S}_z(t) \rangle \right)\gamma_s \Delta t.
\end{equation}
We verify this expression by observing switching between the predicted values from Eq.~\eqref{eq:PDperp}, where $\langle \hat{S}_z(t) \rangle$ is replaced by its mean-field value. This switching occurs at the same time as the transitions in $\hat{S}_z$ shown in Fig.~\ref{fig:quant_jump}(a).

Since the cavity mode is proportional to the collective jump operator $\hat{S}_-$, as follows from Eq.~\eqref{Seq:c_t}, a detection event by the photodetector labeled PD$_{\parallel}$ corresponds to the intensity $\langle \hat{S}_+ \hat{S}_- \rangle$, as shown by the comparison between Fig.~\ref{fig:photon_count}(b) and Fig.~\ref{fig:quant_jump}(e) of the main text. Qualitatively, the number of counts is given by
\begin{equation}\label{eq:PDpar}
    \text{Counts PD}_{\parallel}(t\,|\, t+\Delta t) = \langle \hat{S}_+ \hat{S}_-(t) \rangle \gamma_{\rm c} \Delta t.
\end{equation}
Similar to the individual jumps, the values between switching events in Fig.~\ref{fig:photon_count}(b) are well described by Eq.~\eqref{eq:PDpar}, with $\langle \hat{S}_+ \hat{S}_-(t) \rangle$ replaced by the corresponding expectation values from the states $\rho_\pm$.

The trace of individual jumps is also useful to describe the switching mechanism itself. As can be seen in Fig.~\ref{fig:results} the states $\rho_{\pm}$ have different total momentum $\langle \hat{\bm S}^2 \rangle$, therefore, the switching with $\hat{S}_-$ that conserves $\langle \hat{\bm S}^2 \rangle$ is not possible and individual jumps are required. In Fig.~\ref{fig:short_jump} we show the trajectory with high enough resolution to trace the action of all individual jumps. One can see, that in order to jump from $\rho_+$ to $\rho_-$ many of the individual jumps in the short period of time are required.

\bibliography{new_cit}

\end{document}